\documentclass[twocolumn,twocolappendix,trackchanges]{aastex631}
\usepackage{amsmath}
\usepackage{multirow}
\usepackage{longtable}
\usepackage{graphicx}
\usepackage{appendix}
\usepackage{hyperref}

\received{\today}
\shortauthors{Liu et al.}
\graphicspath{{./}{figures/}}
\usepackage{subfigure}
\usepackage{CJK}
\usepackage{threeparttable}
\usepackage{amsmath}
\usepackage{pifont}
\begin{document}

\title{Investigation of projected rotational velocities of Be-type stars in LAMOST DR7}

\author[0000-0001-5314-2924]{Zhicun Liu}
\affiliation{Department of Physics, Hebei Normal University, Shijiazhuang 050024, People's Republic of China}
\affiliation{Guo Shoujing Institute for Astronomy, Hebei Normal University, Shijiazhuang 050024, People's Republic of China}
\affiliation{Hebei Key Laboratory of Photophysics Research and Application, Shijiazhuang 050024, People's Republic of China}
\affiliation{Shijiazhuang Key Laboratory of Astronomy and Space Science,Shijiazhuang 050024, People's Republic of China}

\author[0000-0002-2577-1990]{Jiao Li}
\affiliation{International Centre of Supernovae (ICESUN), Yunnan Key Laboratory of Supernova Research, Yunnan Observatories, Chinese Academy of Sciences (CAS), Kunming 650216, People's Republic of China}
\affiliation{Key Laboratory for the Structure and Evolution of Celestial Objects, CAS, Kunming 650216, People's Republic of China}

\author[0000-0002-4828-0326]{Jiaming Liu}
\affiliation{Department of Physics, Hebei Normal University, Shijiazhuang 050024, People's Republic of China}
\affiliation{Guo Shoujing Institute for Astronomy, Hebei Normal University, Shijiazhuang 050024, People's Republic of China}

\author[0000-0002-1802-6917]{Chao Liu}
\affiliation{Key Laboratory of Space Astronomy and Technology, National Astronomical Observatories, Chinese Academy of Sciences, Beijing 100101, People's Republic of China}
\affiliation{Institute for Frontiers in Astronomy and Astrophysics, Beijing Normal University, Beijing 102206, People's Republic of China}

\author[0000-0003-2536-3142]{Xiao-Long Wang}
\affiliation{Department of Physics, Hebei Normal University, Shijiazhuang 050024, People's Republic of China}
\affiliation{Guo Shoujing Institute for Astronomy, Hebei Normal University, Shijiazhuang 050024, People's Republic of China}
\affiliation{Shijiazhuang Key Laboratory of Astronomy and Space Science,Shijiazhuang 050024, People's Republic of China}

\author[0000-0003-1828-5318]{Guozhen Hu}
\affiliation{Department of Physics, Hebei Normal University, Shijiazhuang 050024, People's Republic of China}
\affiliation{Guo Shoujing Institute for Astronomy, Hebei Normal University, Shijiazhuang 050024, People's Republic of China}
\affiliation{Shijiazhuang Key Laboratory of Astronomy and Space Science,Shijiazhuang 050024, People's Republic of China}

\author[0000-0003-1359-9908]{Wenyuan Cui}
\affiliation{Department of Physics, Hebei Normal University, Shijiazhuang 050024, People's Republic of China}
\affiliation{Guo Shoujing Institute for Astronomy, Hebei Normal University, Shijiazhuang 050024, People's Republic of China}
\affiliation{Shijiazhuang Key Laboratory of Astronomy and Space Science,Shijiazhuang 050024, People's Republic of China}

\correspondingauthor{Zhicun Liu, Wenyuan Cui}
\email{liuzhicun@hebtu.edu.cn, cuiwenyuan@hebtu.edu.cn}


\begin{abstract}

Stellar rotation plays a key role in the transfer of angular momentum, and a large sample of Be-type stars with reliable projected rotational velocities is crucial for understanding their formation and evolution. In this work, we derive the projected rotational velocities ($v$\,sin\,$i$) of 479 Be-type stars using the Fourier transform method, based on their LAMOST Medium-resolution Survey (MRS) spectra. Our results suggest that the Fourier transform method can provide reliable $v$\,sin\,$i$ values for Be-type stars by analyzing the \ion{He}{1}\,lines at 4922, 5015, 5047, and 6678\,\AA\,in their LAMOST MRS spectra. A K-S test indicates that Be-type stars with different H$\alpha$ emission line morphologies exhibit different $v$\,sin\,$i$ distributions, and Be-type stars with double‑peaked emission have a higher fraction of rapid rotators than those with single-peak emission. The $v$\,sin\,$i$ distributions of our Be-type stars in the field, OB associations, and clusters show no significant differences. The deconvolved $v$\,sin\,$i$ distribution of our entire Be-type star sample does not exhibit a bimodal distribution but rather a single peak at $v\approx260$\,km$\cdot$s$^{-1}$. Based on the analysis of 105 stars in our sample, we find that the mean equatorial rotational velocity is 0.74 times the critical velocity. Furthermore, we investigate the relationship between $v$\,sin\,$i$ and the H$\alpha$ peak separation velocity for Be-type stars exhibiting double-peak H$\alpha$ emission lines, using Pearson and Spearman rank correlation coefficients.
\end{abstract}


\section{Introduction} \label{sec:intro}

Be-type stars, whose spectra show one or more Balmer emission lines, are B-type main sequence stars with luminosity class V to III \citep{1987pbes.coll....3C,1990clst.book.....J,2003PASP..115.1153P}. They have a circumstellar envelope of ionized gas, also called a Keplerian disk, and exhibit excess continuum emission across ultraviolet, optical, and infrared wavelengths \citep{1988A&A...194..167D,1994A&A...290..609D,2015MNRAS.446..274R}. Previous studies indicate that 10-20\% of the B-type stars in the Milky Way are Be-type stars \citep{1983A&A...117..357J,2011BASI...39..517M,2017ApJ...842...48A,Liu2024ApJS..275...24L}. Moreover, Be-type stars also exhibit strong stellar winds, as well as spectral and photometric variations over long timescales that may be caused by their circumstellar disc \citep{2009ssc..book.....G,2012MNRAS.426.2738N}.

Compared to mass and metallicity, stellar rotation also plays an important role in stellar evolution. Observations indicate that Be-type stars have high rotational velocities. \citet{2001A&A...368..912Y} found that the $v$\,sin\,$i$ of late-type Be stars is closer to their critical velocities than that of early-type stars, through a statistical analysis of classical Be stars. \citet{2016A&A...595A.132Z} studied the distribution of rotational velocities of 233 Galactic Be-type stars and found that Be-type stars rotate at about 65\% of their critical velocity. \citet{2022MNRAS.512.3331D} found that Be-type stars in the 30 Doradus region rotate at about 66\% of their critical velocity. Overall, the rotational velocities of most Be-type stars range from 30\% to 95\% of their critical velocities \citep{1996MNRAS.280L..31P,2004PASA...21..310K,2005ApJ...634..585C}. Be-type stars with fast rotation are also ideal for investigating rotational mixing, especially stellar surface nitrogen abundances \citep{2011A&A...536A..65D,2017MNRAS.471.3398A}.

The rapid rotation of Be-type stars is linked to their formation mechanism: single-star evolution or binary interactions. (i) In the single-star evolution scenario, Be-type stars form in two ways. On the one hand, Be-type stars are born with fast rotation and inherit angular momentum from their parent molecular cloud \citep{2008A&A...478..467E,2020A&A...633A.165H,2024A&A...683A..94M}; on the other hand, Be-type stars could originate from B-type stars that may have experienced a spin-up during their main-sequence evolution, transferring their angular momentum from their core to the envelope \citep{1995ARA&A..33..199B,2000A&A...361..101M,2013A&A...553A..25G,2020A&A...641A..42B}. (ii) In the evolution of binary systems, Be-type stars, as the secondary component, could obtain the mass and angular momentum from the primary component via the Roche-lobe, which makes them rapid rotation stars \citep{1991A&A...241..419P,2014ApJ...796...37S,2019ApJ...885..147K,2022MNRAS.516.3602E,2022MNRAS.512.3331D,2020A&A...638A..39L,2021A&A...653A.144H}. The rapid rotation of Be-type stars is one of the main factors for matter ejection, which also leads to their circumstellar disk and emission, but this cannot explain their matter ejection alone, and other mechanisms, such as the beating of nonradial pulsations, are evoked \citep{2003PASP..115.1153P}. 

Studies by \citet{1999A&A...346..459M} and \citet{2002A&A...390..561M} suggest that the ratio of Be-type stars to B-type stars increases as metallicity decreases. \citet{2006ApJ...652..458W} found a higher fraction of Be-type stars in low-metallicity environments, based on their study of the numbers of Be-type stars in the Magellanic Cloud clusters and three Galactic clusters. \citet{2006A&A...452..273M} and \citet{2007A&A...462..683M} found that the Be-type stars in the Magellanic Clouds rotate faster than those in the MW, which is attributed to the weaker stellar winds and less angular momentum loss for the stars formed in lower metallicity regions. Additionally, \citet{2013MNRAS.435.3103I} suggested that the proportion of Be-type stars decreases with increasing cluster age and metallicity by studying the Be-type stars in the Magellanic Clouds. 

The Large Sky Area Multi-Object Fiber Spectroscopic Telescope (LAMOST), also called the Guoshoujing Telescope, is a 4\,m reflective Schmidt telescope with 4000 fibers \citep{2006ChJAA...6..265Z,2012RAA....12.1197C,2012RAA....12..723Z}. These features make LAMOST efficient in obtaining spectra. Furthermore, the Medium Resolution Survey (MRS) of LAMOST began in 2018, and the LAMOST MRS spectra with $R\sim7500$ are made with the blue band (4950--5350\,\AA) and red band spectra (6300--6800\,\AA) \citep{2020arXiv200507210L}. In addition, there are some LAMOST MRS spectra including the \ion{He}{1} $\lambda$4922 absorption line. The LAMOST DR7 dataset includes 9,846,793 stellar spectra with $R\sim1800$ and 11,190,570 single-exposure spectra with $R\sim7500$ \footnote{http://dr7.lamost.org/}. 

In this paper, we aim to derive $v$\,sin\,$i$ values and explore the distribution of $v$\,sin\,$i$ for Be-type stars in LAMOST DR7 using their MRS spectra and the Fourier transform method. In Section \ref{sec:Data}, we describe the process of constructing the sample. The Fourier transform method is described in Section \ref{sec:Method}. The results and discussion are given in Section \ref{subsec:Results}, while the conclusions are summarized in Section \ref{sec:Conclusions}.  

\section{sample data} \label{sec:Data}

In this study, we compiled a large sample of Be-type stars from LAMOST DR7 MRS spectra. This sample includes 479 Be-type stars selected from three different catalogs. The procedure is described as follows.

(i) \citet{Wang2022ApJS..260...35W} presented a catalog of Be-type stars from the LAMOST DR7 MRS, based on a deep convolutional neural network (ResNet\footnote{For more information, see \citet{Wang2022ApJS..260...35W}}). This catalog contains 334 classical Be-type stars and 312 Be-type star candidates showing shell line features in the H$\alpha$ profile. 

(ii) \citet{Guo2022RAA....22b5009G} identified 9382 early-type stars with S/N$>$40, measuring the equivalent widths of several spectral lines (H$\alpha$ 6564\,\AA, \ion{He}{1} 6678\,\AA, and \ion{Mg}{1} profiles in the range of 5167$-$5184\,\AA). 

\begin{figure*}[!htp]
    \centering
    \includegraphics[width=1.0\linewidth]{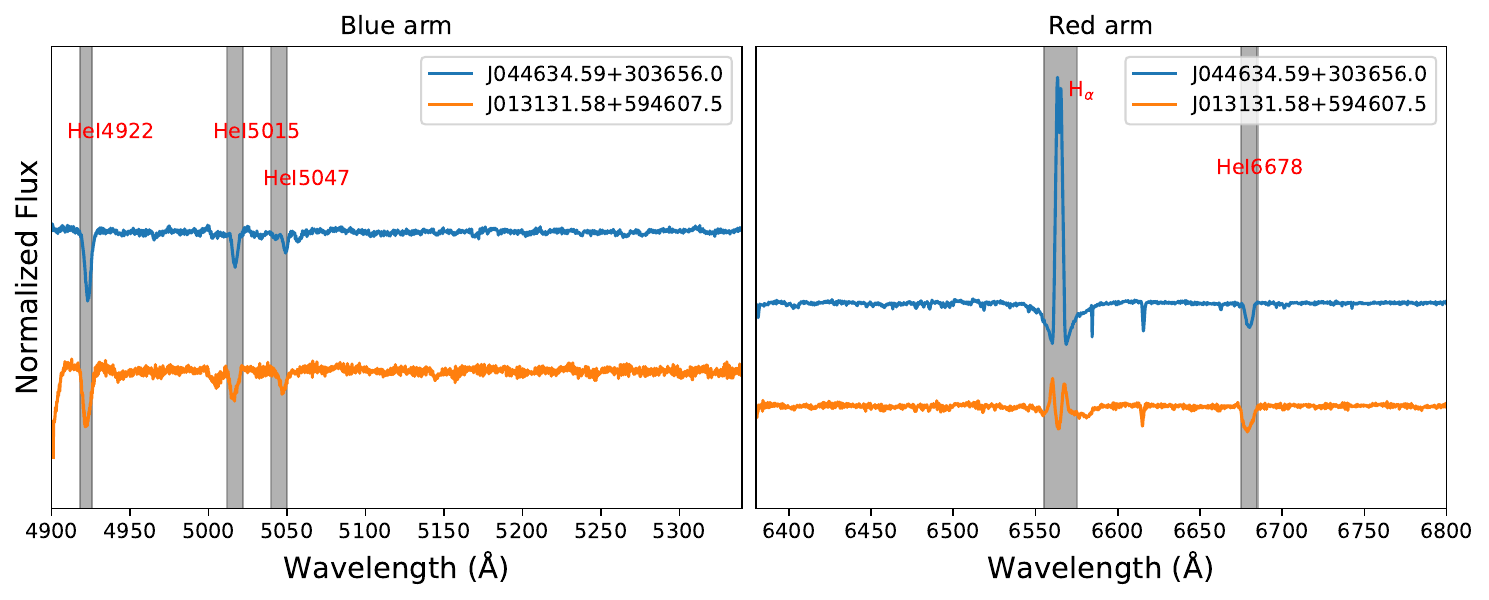}
    \caption{The LAMOST continuum-normalized blue- (left panel) and red-arm (right panel) medium-resolution spectra of two Be-type stars (J044634.59+303656.0 and J013131.58+594607.5) in our sample. Some important line features, such as \ion{He}{1}\,4922, 5015, 5047, 6678\,\AA, and H$_\alpha$ are also marked.}
    \label{fig00}
\end{figure*}

\begin{figure}[htp]
    \centering
    \includegraphics[width=1.0\linewidth]{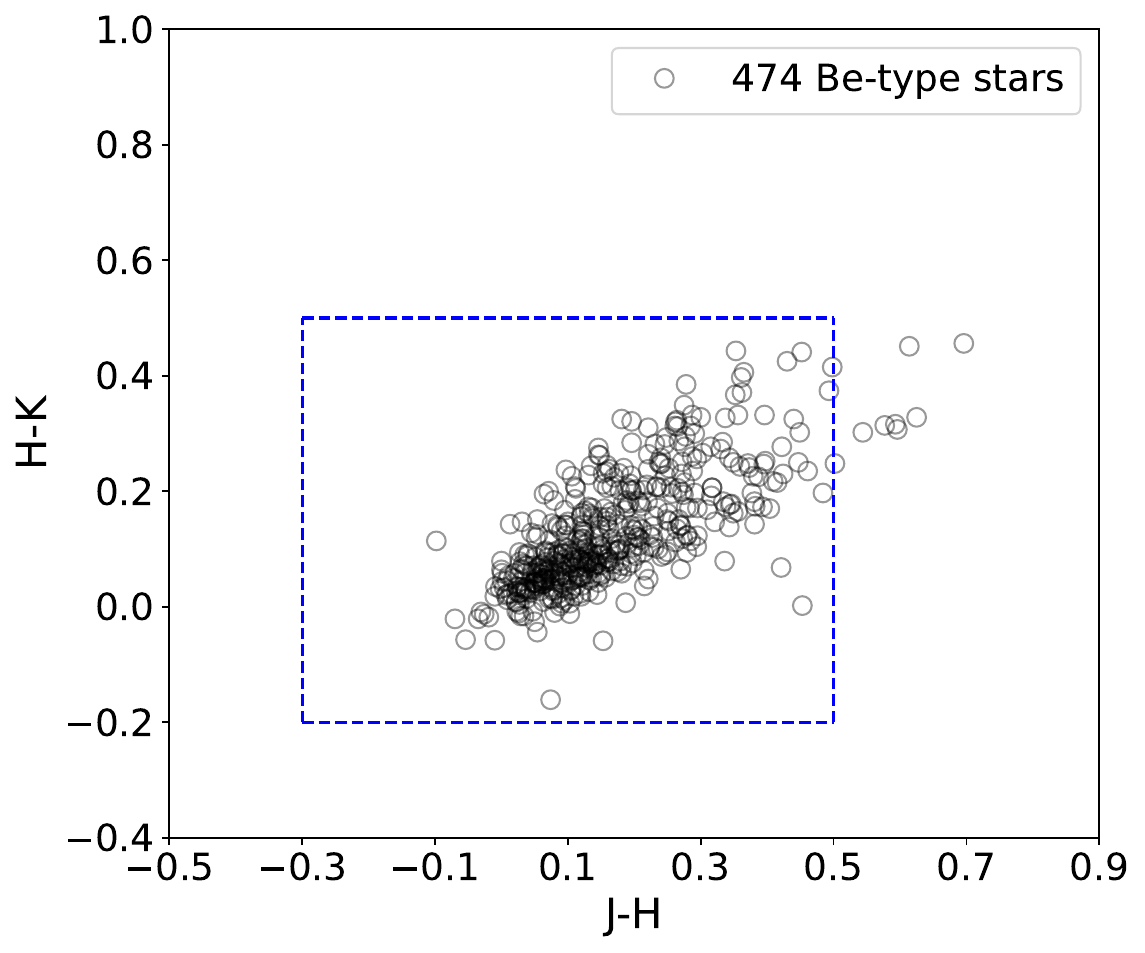}
    \caption{The J-H versus H-K color-color diagram for 474 of 479 Be-type stars in this work. The Be-type stars in the blue area correspond to the area $-$\,0.2$<$\,(H–K)\,$<$\,0.5 and $-$\,0.3 $<$ (J–H)\,$<$\,0.5 defined by \citet{2016MNRAS.463.1162C}.}
    \label{fig01}
\end{figure}

(iii) Recently, \citet{Liu2024ApJS..275...24L} presented a value-added catalog of OB stars in the LAMOST DR7 low-resolution spectroscopic survey using a modified method of identifying OB stars \citep{Liu2019ApJS..241...32L}. This catalog includes 3006 Be-type stars and candidates. We cross-matched this catalog with the LAMOST DR7 MRS catalog and obtained 426 common stars.

To derive the $v$\,sin\,$i$ values of Be-type stars using the Fourier transform method, we selected the Be-type stars whose spectra have a sufficient signal-to-noise ratio (S/N$\geq$40), H$\alpha$ emission, and clear helium absorption lines at 4922, 5015, 5047, and 6678\,\AA\, as our sample. The selection criteria for a spectral signal-to-noise ratio S/N$\geq$40 is described in Appendix~\ref{appendb}. We examined the LAMOST medium-resolution spectra of these stars from the three catalogs and identified 479 Be-type stars as our final sample \footnote{During the sample selection, we chose the highest-S/N spectra for the Be-type stars that had multiple observations.}. In addition, we exclude the double-lined spectroscopic binaries. Figure~\ref{fig00} presents the LAMOST medium-resolution spectra of two representative Be-type stars from the final sample.

\begin{figure*}[htp!]
    \centering
    \includegraphics[width=14cm,height=7cm]{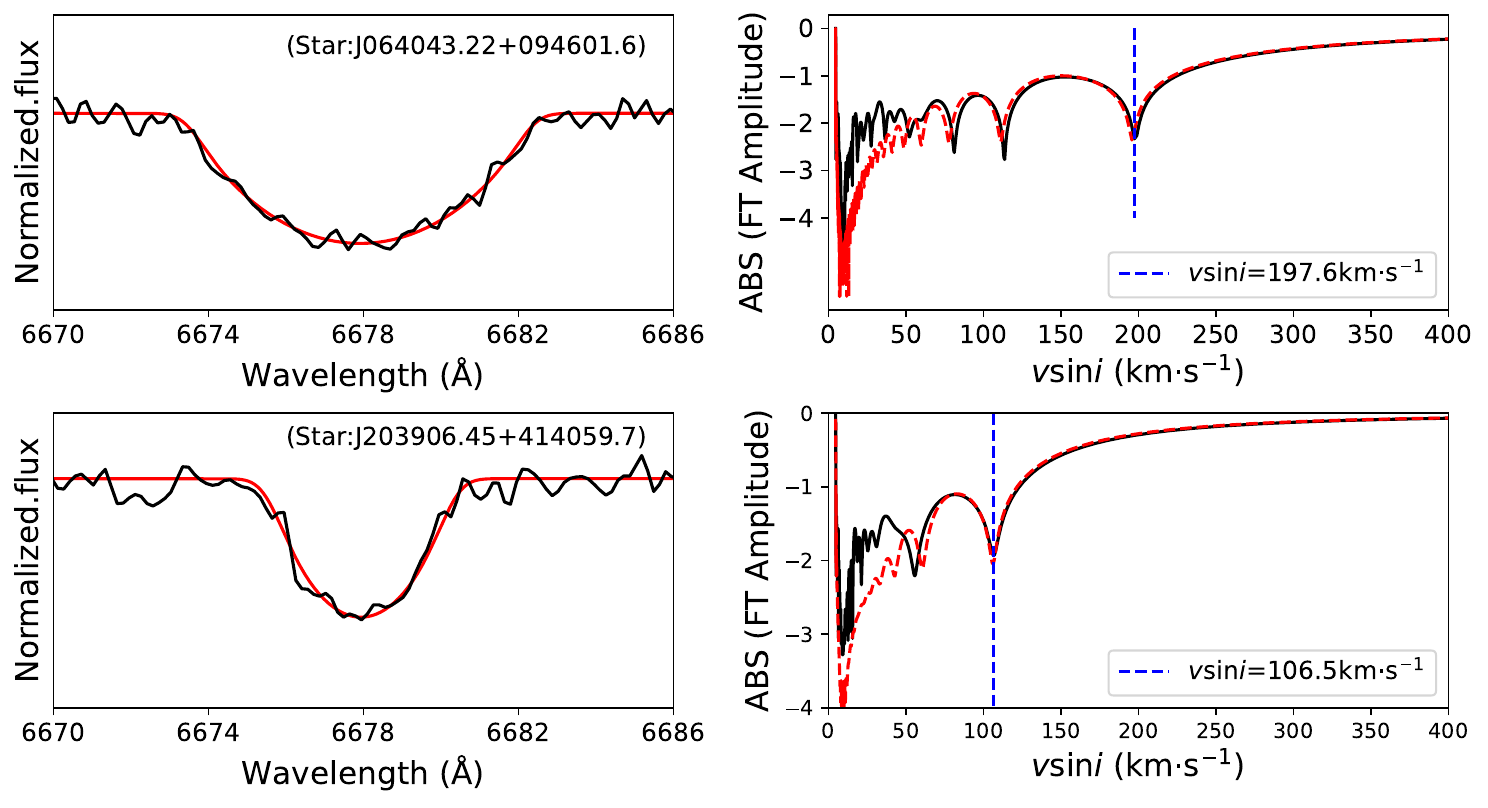}
    \caption{Left: observed spectral line profiles (solid black line) and theoretical rotational profiles (solid red line) of the \ion{He}{1} 6678\,\AA\, line for J064043.22+094601.6 (top; S/N$_{\rm R}$=382.33) and J203906.45+414059.7 (bottom; S/N$_{\rm R}$=62.68). Right: Fourier transform of the observed profiles (solid black line) and the theoretical rotational profiles (dashed red lines) for J064043.22+094601.6 with $v$\,sin\,$i$ = 197.6\,km$\cdot$s$^{-1}$ (top) and J203906.45+414059.7 with $v$\,sin\,$i$=106.5\,km$\cdot$s$^{-1}$ (bottom). The position of the first minimum in the FT is marked by a vertical dashed blue line.}
    \label{fig02}
\end{figure*}

\begin{table*}[htp!]
\footnotesize
\centering
\caption{Information for the 479 Be-type stars in this study. Columns from left to right are: Obsid, Designation, the spectral signal-to-noise ratios in the blue (S/N$_{\rm B}$) and red (S/N$_{\rm R}$) bands from LAMOST, SIMBAD designation, $v$\,sin\,$i$ values derived from individual \ion{He}{1} lines, the mean value of $v$\,sin\,$i$ ($\overline{v\,{\rm sin}\,i}$), the uncertainty of $v$\,sin\,$i$ ($\delta$\,$v$\,sin\,$i$), the velocity separation of the double peaked profiles of H$\alpha$ ($\Delta$H$\alpha$), the H$\alpha$ morphology (MH$\alpha$), and the membership classification (MC). For Be-type stars with $v$\,sin\,$i$$\leq$100, the $\delta$\,$v$\,sin\,$i$ is set to 10\,km$\cdot$s$^{-1}$, and for Be-type stars with $v$\,sin\,$i$$>$100\,km$\cdot$s$^{-1}$, the $\delta$\,$v$\,sin\,$i$ is 10\% of $\overline{v\,{\rm sin}\,i}$. This full Table is available in FITS format.}\label{Table1}
\begin{tabular}{llrrllllllllcl}
\hline
  \multicolumn{1}{c}{Obsid} &
  \multicolumn{1}{c}{Designation} &
  \multicolumn{2}{ c}{S/N (LAMOST)} &
  \multicolumn{1}{c}{Designation} &
  \multicolumn{4}{c}{$v$\,sin\,$i$ (km$\cdot$s$^{-1}$)} &
  \multicolumn{1}{l}{$\overline{v\,{\rm sin}\,i}$} &
  \multicolumn{1}{l}{$\delta$\,$v$\,sin\,$i$} &
  \multicolumn{1}{l}{$\Delta$H$\alpha$} &
  \multicolumn{1}{l}{MH$\alpha$} &
  \multicolumn{1}{l}{MC} \\
  \cline{3-4}
  \cline{6-9}
  \cline{10-12}
  \multicolumn{1}{l}{(LAMOST)} &
  \multicolumn{1}{c}{(LAMOST)} &
  \multicolumn{1}{c}{B} &
  \multicolumn{1}{c}{R} &
  \multicolumn{1}{c}{(SIMBAD)} &
  \multicolumn{1}{c}{4922\AA} &
  \multicolumn{1}{c}{5015\AA} &
  \multicolumn{1}{c}{5047\AA} &
  \multicolumn{1}{c}{6678\AA} &
  \multicolumn{3}{c}{(km$\cdot$s$^{-1}$)} &
  \multicolumn{1}{c}{(A)} &
  \multicolumn{1}{c}{(B)}\\
\hline
591005055& J205356.93+500529.4&181.60&272.44&LS\,III\,+49\,11&\nodata & \nodata & 345.2  &320.6  &332.9&33.3 &\nodata&b &FD\\
591509201& J012318.25+580435.1&101.11& 140.87&BD+57\,273&\nodata & \nodata & 394.9  &\nodata& 394.9&39.5& 216.7&b&OC\\
591515181& J011524.01+583108.5&113.75& 138.13&BD+57\,230&\nodata & \nodata &\nodata & 284.6 & 284.6&28.5& \nodata&c&OC\\
591608032 &J012319.41+573851.7&212.22& 285.44&BD+56\,259 &\nodata & 147.3 &\nodata  & 164.6 & 156.0 & 15.6&\nodata  & c & FD\\
591615105 &J011902.35+581920.2&86.67 & 111.76&NGC\,457\,124&\nodata  &\nodata  &\nodata  & 280.5 & 280.5 &28.0& 189.9 & b & OC\\
\nodata & \nodata  \nodata & \nodata & \nodata & \nodata & \nodata & \nodata & \nodata & \nodata& \nodata& \nodata& \nodata& \nodata \\
\nodata & \nodata  \nodata & \nodata & \nodata & \nodata & \nodata & \nodata & \nodata & \nodata& \nodata& \nodata& \nodata  & \nodata\\
\nodata & \nodata  \nodata & \nodata & \nodata & \nodata & \nodata & \nodata & \nodata & \nodata  & \nodata& \nodata& \nodata& \nodata\\
\hline
\multicolumn{14}{l}{(A): The a, b,c ,d, represent the different morphologies of H$\alpha$, respectively (a for Be-type stars with a single emission peak, b for Be-type stars } \\ 
\multicolumn{14}{l}{\quad \quad with double‑peak emission, c for Be-type stars with a narrow emission peak superimposed on top of an absorption profile, and d for Be- }\\
\multicolumn{14}{l}{\quad \quad type stars with an absorption shell, seeing Figure \ref{fig05}.} \\
\multicolumn{14}{l}{(B) MC represents the membership classification (FD for Be-type stars in the field, OC for Be-type stars in the open clusters, and OS for }\\
\multicolumn{14}{l}{\quad \quad Be-type stars in the OB associations.)}\\
\end{tabular}
\label{tab:my_label}
\end{table*}

We cross-matched the 479 Be-type stars with $\rm {2MASS}$ to obtain their photometry in the J, H, and K bands. Figure~\ref{fig01} shows the distribution of 474 Be-type stars in the (J–H) versus (H–K) diagram. Most fall within the region $-$\,0.2$<$\,(H–K)\,$<$\,0.5 and $-$\,0.3 $<$ (J–H)\,$<$\,0.5, which is characteristic of classical Be stars \citep{2016MNRAS.463.1162C}, confirming that our sample consists of genuine Be stars.

\section{projected rotational velocity} \label{sec:Method}

\subsection{Methodology}

There are four main methods used to estimate the projected rotational velocities of OB-type stars, as follows: (i) Full-width at half-maximum (FWHM): A direct measure of the FWHM of the spectral lines to derive $v$\,sin\,$i$ \citep[see, for example][]{2002ApJ...573..359A,2005AJ....129..809S,2012AJ....144..130B,2015AJ....150...41G}). (ii) Cross-correlation (CC): The Gaussian width of the cross-correlation function, which is calculated by using the observed spectrum and template spectra with different rotational velocities, is directly related to the stellar $v$\,sin\,$i$ \citep{1996ApJ...463..737P,1997MNRAS.284..265H}. (iii) Profile fitting (PF): Comparing observed line profiles with theoretical counterparts convolved with a rotational broadening function \citep[see, for example][]{1992oasp.book.....G,2006A&A...457..265D,2008A&A...479..541H}. (iv) Fourier transform (FT): The position of the zeroes of the FT depends on the value of $v$\,sin\,$i$ (see, for example, \citep{2013A&A...550A.109D,2007A&A...468.1063S,2014A&A...562A.135S,2022MNRAS.512.3331D}). 

\begin{figure}[htp!]
    \centering
    \includegraphics[width=1.0\linewidth]{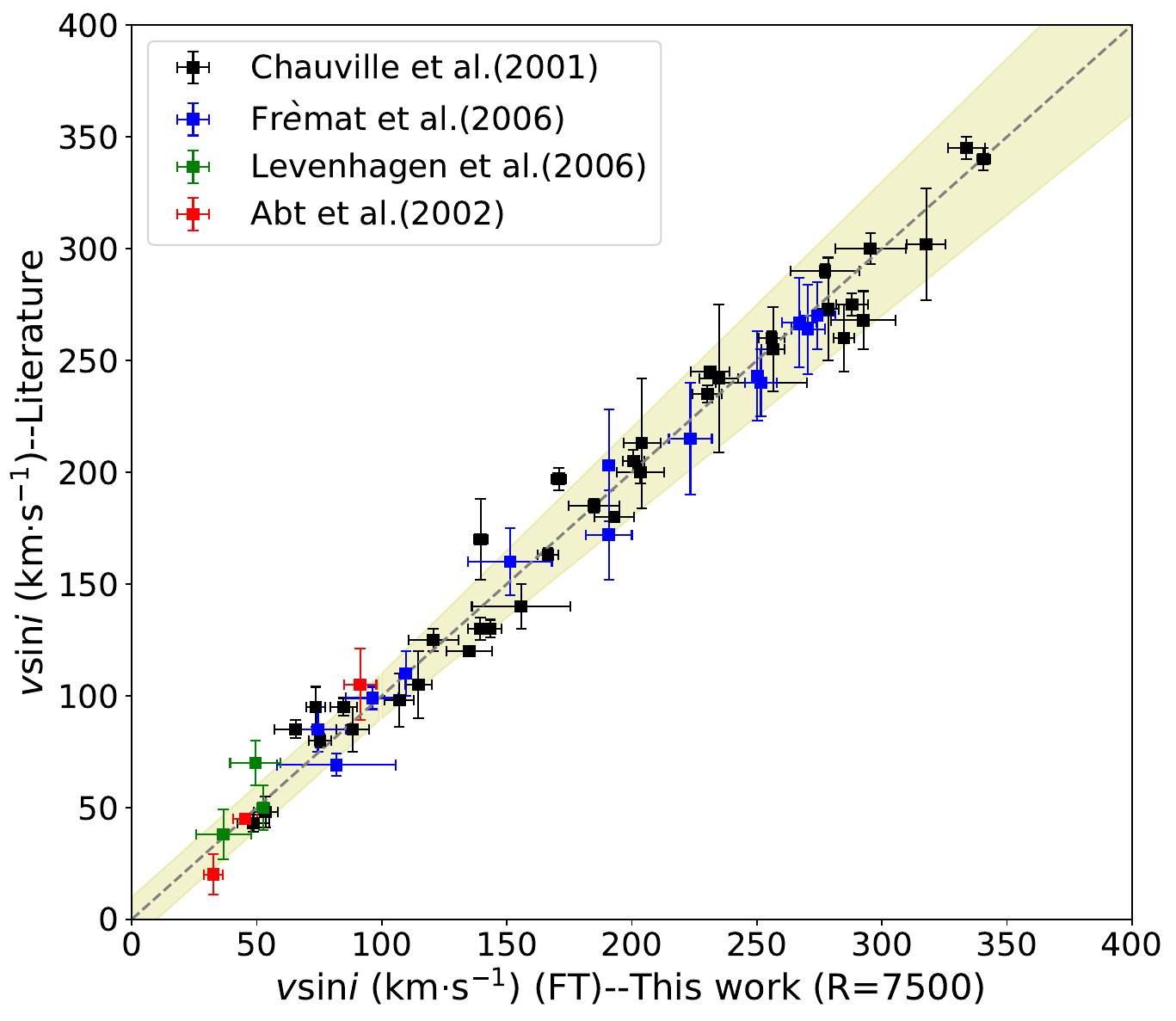}
    \caption{Comparison of $v$\,sin\,$i$ values derived using the FT method and those from the literature for 56 Be-type stars. Different color symbols represent sources from different literature. The gray dashed line represents a one-to-one relationship. The yellow regions indicate a 10\,\% or 10\,km$\cdot$s$^{-1}$ uncertainty, whichever is larger.}
    \label{fig03}
\end{figure}

Each of these methods has its advantages and disadvantages. FWHM and CC are directly related to $v$\,sin\,$i$, but they do not take into account the effect from non-rotational broadening, such as Stark broadening of the H and He lines, and macroturbulence \citep{2002MNRAS.336..577R,2008A&A...479..541H,2014A&A...562A.135S}. PF can estimate the rotational velocity of a star with lower-quality spectra, but it relies on the intrinsic line profile from a stellar atmosphere model and on the broadening mechanisms that affect the line profile. FT can separate rotational broadening from other broadening mechanisms, such as macroturbulence, although it needs high-quality spectra. Additionally, \citet{2007A&A...468.1063S,2014A&A...562A.135S} studied the use of FT on early-type stars and indicated that macroturbulence is common in massive stars, especially in OBA-type stars.

The first minimum in the FT for a spectral line relies on the given spectral line, and it is assumed that the first zero in the FT is only related to the rotational broadening. \citet{2007A&A...468.1063S} investigated the $v$\,sin\,$i$ of Galactic OB-type stars using the FT method. They found that this method can reliably derive the projected rotational velocities for OB-type stars by using \ion{He}{1} or metal lines (for elements such as O, Si, Mg). Furthermore, they noted that the absorption lines at different wavelengths can give different values of the minimum $v$\,sin\,$i$. 

Considering the wavelength range of LAMOST MRS spectra and the range of projected rotational velocities of Be-type stars from the literature, we adopted the FT method to estimate the $v$\,sin\,$i$ values of 479 Be-type stars, based on the \ion{He}{1}\,4922, 5015, 5047, and 6678\,\AA\,lines. The $v$\,sin\,$i$ results are provided in Table~\ref{Table1}. The left panel of Figure~\ref{fig02} presents a comparison between the observed spectra and the FT spectra at the \ion{He}{1}\,6678\,\AA for J064043.22+094601.6 (S/N$_{\rm R}$=382.33) and for J203906.45+414059.7 (S/N$_{\rm R}$=62.68). The right panel of Figure \ref{fig02} presents the Fourier transform of the observed profile (solid black line) and of the theoretical rotational profile (dashed red line) with $v$\,sin\,$i$= 197.6 km\,$\cdot$\,s$^{-1}$ and $v$\,sin\,$i$= 106.5 km\,$\cdot$\,s$^{-1}$. The central vertical dashed blue line shows the position of the first zero.

\subsection{Validation}

To validate the reliability of $v$\,sin\,$i$ for Be-type stars derived with the FT method using medium-resolution spectra, we collected 56 Be-type stars from \citet{Chauville2001A&A...378..861C}, \citet{Neiner2006A&A...451.1053F}, \citet{Levenhagen2006MNRAS.371..252L}, and \citet{2002ApJ...573..359A}. These stars have $v$\,sin\,$i$ values ranging from 20 to 350\,km\,$\cdot$\,s$^{-1}$, and serve as our validation sample. We obtained their high-resolution spectra \footnote{High-resolution spectra are obtained from \url{http://basebe.obspm.fr/basebe/}.} and reduced them to match the resolution of LAMOST MRS ($R \sim 7500$). We then measured their $v$\,sin\,$i$ values using the FT method, based on the four different helium absorption lines at different resolutions (where possible). 

\begin{figure}[t!]
    \centering
    \includegraphics[width=0.99\linewidth]{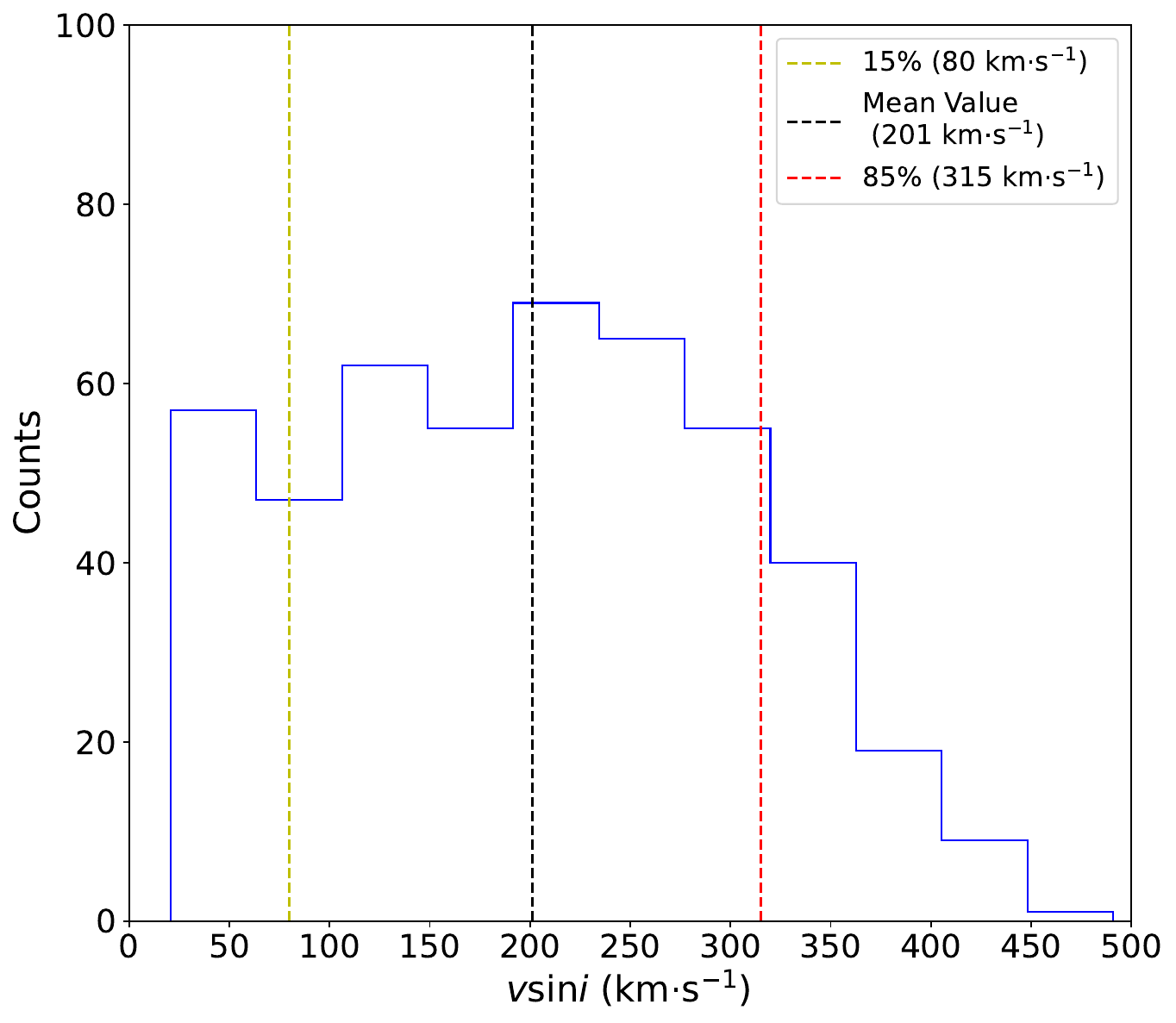}
    \caption{The $v$\,sin\,$i$ distribution of 479 Be-type stars. The yellow, black and red dashed lines represent the 15th percentile (80\,km$\cdot$s$^{-1}$), the mean value (201\,km$\cdot$s$^{-1}$) and the 85th percentile (315\,km$\cdot$s$^{-1}$) of $v$\,sin\,$i$, respectively.}\label{fig04} 
\end{figure}

\begin{figure*}[htp!]
    \centering
    \includegraphics[width=16cm,height=10cm]{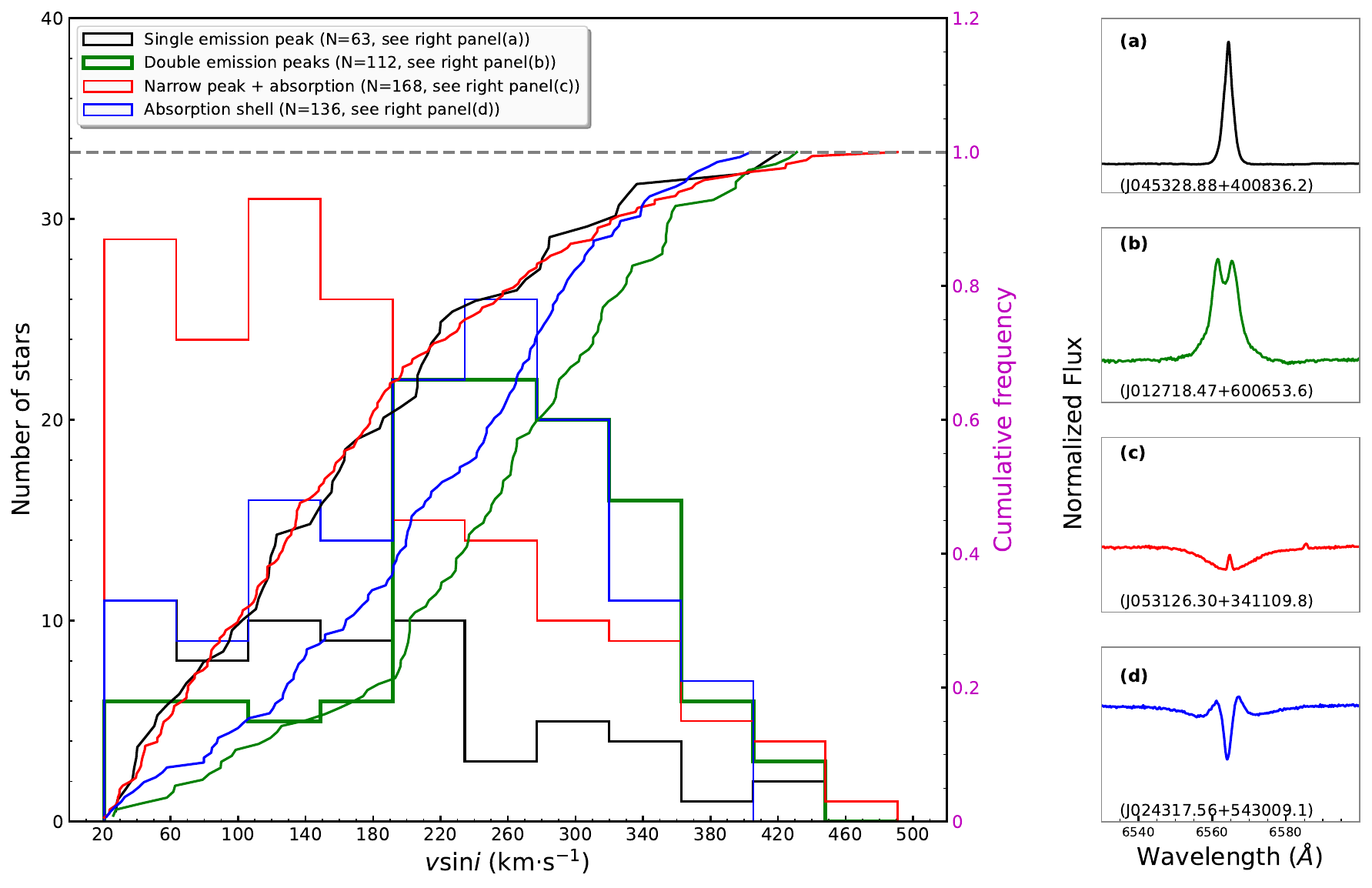}
    \caption{Left: histogram (left y-axis) and cumulative fractions (right y-axis) of $v$\,sin\,$i$ for 479 Be-type stars with different H$\alpha$ emission line morphologies. The number of Be-type stars in each category is also marked. Right: from top to bottom, the LAMOST MRS spectra show H$\alpha$ emission profiles of four Be-type stars, each with a different H$\alpha$ emission line morphology. The colors used for the histogram and cumulative fractions in the left panel match those of the corresponding H$\alpha$ emission lines shown in the right panel.}
\label{fig05}
\end{figure*}

Figure \ref{fig03} compares the $v$\,sin\,$i$ values from this study (using reduced spectra, $R\sim7500$) with those reported in the literature (from high-resolution spectra) for the 56 Be-type stars. The $v$\,sin\,$i$ values for these stars in this work are obtained using the mean value of four \ion{He}{1} lines (where possible), with the errors corresponding to the standard deviation. The good agreement confirms the reliability of our FT-based method for Be-type stars with $v$\,sin\,$i$$>$20\,km$\cdot$s$^{-1}$. We note that only three stars in the validation sample have $v$\,sin\,$i$$<$50\,km$\cdot$s$^{-1}$. Taking into account of the uncertainties in $v$\,sin\,$i$ derived from different helium lines and from the validation sample, we adopt the following empirical uncertainties: 10\,km$\cdot$s$^{-1}$ for stars with $v$\,sin\,$i$$\leq$100\,km$\cdot$s$^{-1}$, and an uncertainty of 10\% for stars with $v$\,sin\,$i$$>$100\,km$\cdot$s$^{-1}$. Given the small validation sample at low rotational velocities, user should be cautious when using Be-type star data with $v$\,sin\,$i$$<$50\,km$\cdot$s$^{-1}$.

\section{Results and discussions} \label{subsec:Results}
\subsection{The entire Be-type star sample}

Figure \ref{fig04} displays the $v$\,sin\,$i$ distribution of 479 Be-type stars. The mean value of $v$\,sin\,$i$ of Be-type stars in this study is 201\,km$\cdot$s$^{-1}$, and most stars have high $v$\,sin\,$i$ values. 
The overall distribution appears broadly unbimodal, but there is an excess of stars at low $v$\,sin\,$i$ $<$150\,km$\cdot$s$^{-1}$ and a sharp decline beyond 300\,km$\cdot$s$^{-1}$. This distribution, a main concentration around 200–280\,km$\cdot$s$^{-1}$ together with a small low-velocity component and a high-velocity cutoff, supports the idea that Be-type stars are produced by two mechanisms: single-star evolution and binary system evolution \citep{2020A&A...633A.165H,2024A&A...683A..94M,2014ApJ...796...37S,2020A&A...638A..39L}. In addition, binary interactions, magnetic fields, and sample selection effects also influence the $v$\,sin\,$i$ distribution of Be-type stars \citep{2012Sci...337..444S,2015AJ....150...41G}. 

\begin{figure*}[htp!]
    \centering
    \includegraphics[width=0.75\linewidth]{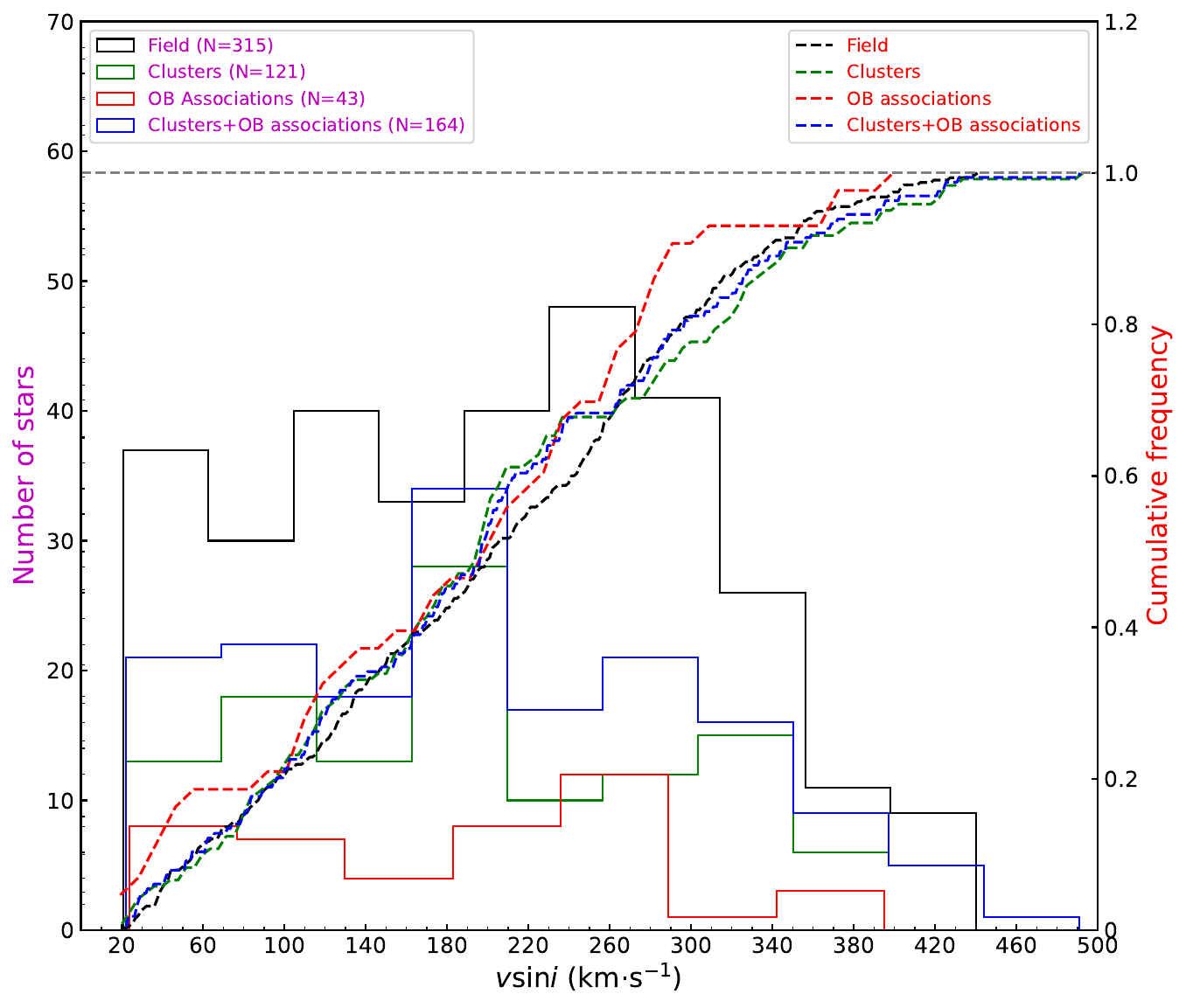}
    \caption{The histogram (left Y-axis) and cumulative fractions (right Y-axis) of $v$\,sin\,$i$ for Be-type stars in the field (black), clusters (green), OB associations (red), and the combined sample of clusters and OB associations (blue). The gray dashed line indicates where the cumulative fractions equal one.}
    \label{fig06}
\end{figure*}

Figure~\ref{fig05} shows the histogram and the cumulative fractions of $v$\,sin\,$i$ for Be-type stars with different H$\alpha$ emission line morphologies. Among the four subtypes, Be-type stars with double‑peak emission (green) have the highest $v$\,sin\,$i$ values, peaking at 200–280\,km$\cdot$s$^{-1}$, indicating the highest fraction of fast rotators. Be-type stars with an absorption shell (blue), also called B-type shell stars, show intermediate $v$\,sin\,$i$ values, peaking at 240–280\,km$\cdot$s$^{-1}$, corresponding to an intermediate fraction of fast rotators. In contrast, both Be-type stars with single‑peak emission (black) and those with narrow-peak emission with absorption (red) display the lowest $v$\,sin\,$i$ values, reflecting the lowest fraction of fast rotators and similar rotational velocity distributions.

\begin{table}[ht!]
    \centering
    \caption{Summary of the K‑S test results for Be-type stars with different H$_\alpha$ emission line morphologies.}
    \begin{tabular}{lll}
    \hline
    Be-type stars with different   & $p$-value& Significance \\
    H$_\alpha$ emission line morphologies&&($\alpha$= 5$\times$10$^{-2}$)\\
    \hline
    Single peak vs Double peaks&3.13×10$^{-6}$ &Significant\\
    Single peak vs Narrow+Abs$^a$& 9.61×10$^{-1}$&Not significant\\
    Single peak vs B-type shell&4.08×10$^{-3}$ &Significant\\
    Double peaks vs Narrow+Abs&1.06×10$^{-12}$ &Significant\\
    Double peaks vs B-type shell&2.82×10$^{-2}$ &Significant\\
    Narrow+Abs vs B-type shell&2.27×10$^{-6}$&Significant\\
    \hline
    \multicolumn{3}{l}{$^a$ Narrow+Abs represents the Be-type stars with} \\
    \multicolumn{3}{l}{\quad narrow‑peak emission with absorption.} \\
    \end{tabular}
    \label{Table2}
\end{table}

To statistically evaluate these differences, we performed two‑sample Kolmogorov‑Smirnov (K‑S) tests and present the results in Table~\ref{Table2}. Except for the comparison between Be-type stars with single‑peak emission and those with narrow‑peak emission with absorption, all other morphological pairs show significant differences ($p<$0.05). Be-type stars with double‑peak emission differ significantly from Be-type stars with single-peak emission and from those with narrow‑peak emission with absorption. Be-type shell stars also differ significantly from all the other three subtypes. These results suggest that the $v$\,sin\,$i$ distributions are systematically different across H$_\alpha$ morphological types.

\subsection{Stellar environment and rotational velocity}

Previous studies suggest that the rotational velocities of early-type stars in different stellar environments exhibit different distributions \citep{2006A&A...452..273M,2007A&A...462..683M,2015AJ....150...41G}. B-type stars formed in high-density regions have higher rotational velocities than those formed in low-density regions \citep{2007AJ....133.1092W}. B-type stars in cluster show a larger mean $v$\,sin\,$i$ than those in the field \citep{2006ApJ...648..580H,2008ApJ...683.1045H}. We cross-matched our Be-type star sample with open clusters and OB associations from the literature (open cluster: \citep{2018A&A...618A..93C,2020A&A...635A..45C,2022A&A...661A.118C,2023ApJS..266...36C,2022ApJS..260....8H,2022ApJS..262....7H,2023ApJS..264....8H,2023ApJS..267...34H,2021ApJ...923..129J,2021A&A...646A.104H,2023A&A...673A.114H}, and OB associations: \citep{Chemel2022MNRAS.515.4359C,liu2025ApJS..280...28L}. We found 43 Be-type stars in OB associations,  and 121 in clusters, and classified the remaining 315 Be-type stars as field stars. Compared to B-type stars, Be-type stars have a higher field fraction \citep{Dallas2022ApJ...936..112D}.

The histogram and cumulative fractions of $v$\,sin\,$i$ for Be-type stars in clusters, OB associations, and fields are also displayed in Figure \ref{fig06}. It can be seen that, unlike in OB associations, a small number of Be-type stars in clusters and the field are great than 400\,km$\cdot$s$^{-1}$ in $v$\,sin\,$i$. The overall $v$\,sin\,$i$ distributions of the three groups are similar, and most stars are located between 20 and 360\,km$\cdot$s$^{-1}$. Be-type stars in OB associations rotate more slowly than those in the field. In fact, this discussion relates to whether OB stars can form in isolation and whether all OB stars, even if now isolated, once belonged to a cluster \citep{2012AJ....144..130B}. If Be-type stars form in clusters, they would have higher rotational velocities due to the high-density gas in clusters \citep{2007AJ....133.1092W,Wolff2008AJ....136.1049W}. OB associations, as the medium stage between cluster stars and field stars, are \textbf{expected} to have rotational velocities larger than those of field stars but lower than those of cluster stars \citep{Lada2003ARA&A..41...57L,Wright2020NewAR..9001549W}. Recent studies have indicated that star formation occurs over a continuum of densities rather than at either low or high densities \citep{Bressert2010MNRAS.409L..54B,Kerr2021ApJ...917...23K}. Compared to Be-type stars in clusters and the fields, those in OB associations have a higher fraction of slowly rotating stars. This might be due to the following two main reasons: (i) the OB associations identified by \citet{liu2025ApJS..280...28L} and \citet{Chemel2022MNRAS.515.4359C} are looser and larger in size, implying that the Be-type stars in this study may have formed in low-density regions. (ii) the sample size of Be-type stars is relatively small, which may introduce a selection effect. 

The distribution of $v$\,sin\,$i$ for Be-type stars in the field appears to show a bimodal distribution, which may imply that Be-type stars in the field with different $v$\,sin\,$i$ have different origins. The cumulative distributions of $v$\,sin\,$i$ of Be-type stars in the field, clusters, and OB associations show slight differences. To statistically compare the $v$\,sin\,$i$ distributions of Be-type stars in different environments, we performed two-sample K‑S tests. The results reveal no significant differences between any pair of environments: field versus clusters ($p$=0.54), field versus OB associations ($p$=0.39), and field versus the combined sample of OB associations and clusters ($p$=0.33). The consistently high $p$-values (all $>0.3$) indicate no statistically significant differences in the $v$\,sin\,$i$ distributions of Be-type stars across different stellar environments.

\subsection{Deconvolution of the $v$\,sin\,$i$ distribution in the Be-type star sample}

\begin{figure}[htp!]
    \centering
    \includegraphics[width=1.0\linewidth]{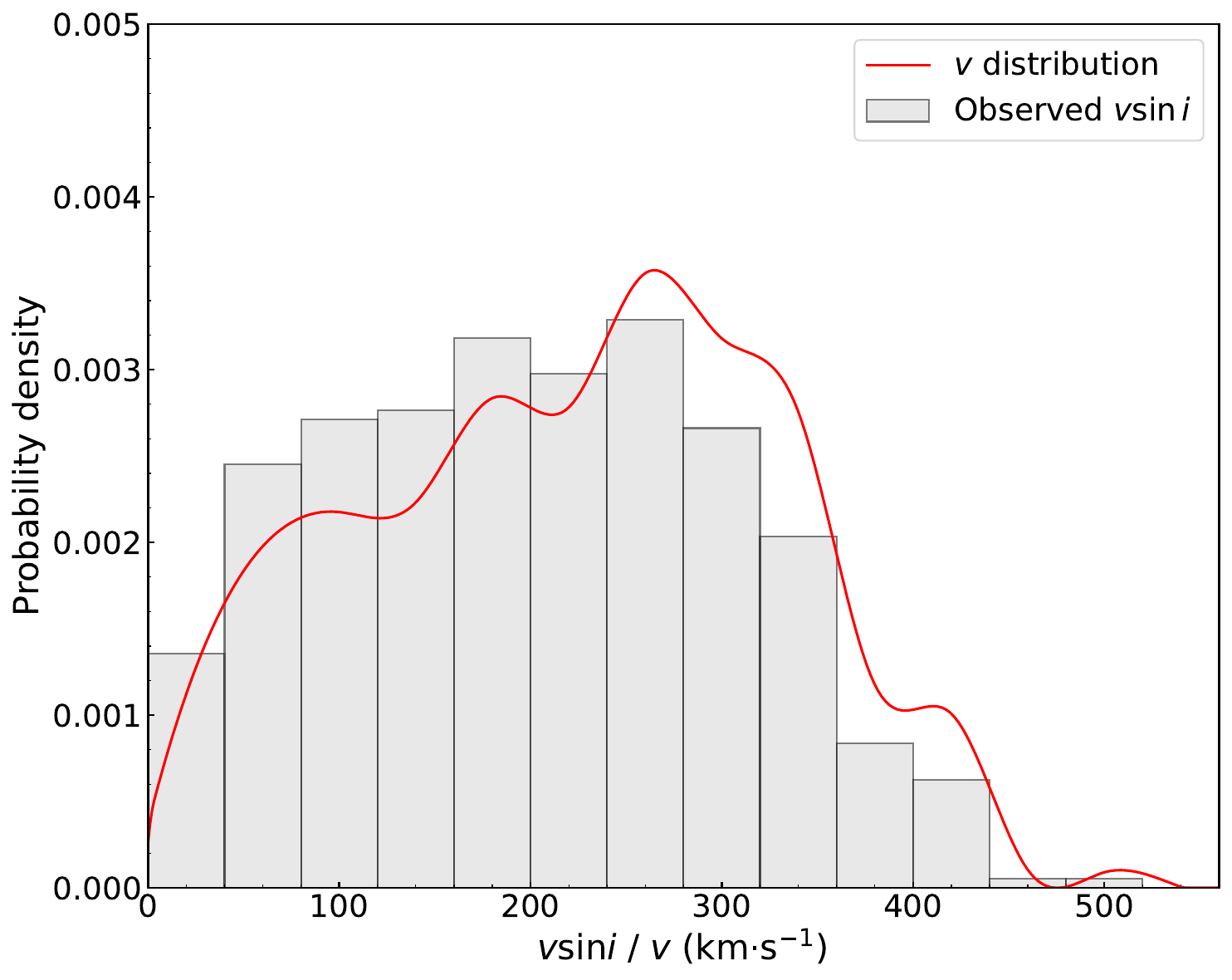}
     \caption{The distribution of rotation velocities of 479 Be-type stars. The black histogram represents the observed $v$\,sin$i$ distribution binned to 40\,km$\cdot$s$^{-1}$. The red solid line marks the deconvolved $v$\,sin$i$ ($v$), which was obtained using the iterative procedure of \citet{1974AJ.....79..745L}, assuming a Gaussian distribution and using four iterations.}
    \label{fig007}
\end{figure}

$v$\,sin\,$i$ is the stellar projected rotational velocity. To calculate the equatorial rotation velocity ($v$), we need to know the inclination angle ($i$). The stellar inclination angles can be determined using interferometric techniques \citep{2004A&A...418..781D,2014A&A...569A..10D,2011A&A...526A..87Z,2025A&A...695A.129T}. We can infer the probability density for the rotational velocity distribution ($P_v$) from that for $v$\,sin\,$i$ under the assumption of random rotation axes. This method has been used to optimize the fit to observations after convolution to allow for random inclination, based on different analytical functions \citep{2006A&A...456.1131M,2006A&A...457..265D,2008A&A...479..541H,2013A&A...550A.109D,2023A&A...680A..32B}. To estimate the distribution of $v$, we adopted the iterative procedure of \citet{1974AJ.....79..745L} assuming a Gaussian distribution and using four iterations.

As shown in Figure~\ref{fig007}, the deconvolved rotational velocity distribution of our entire sample of Be-type stars does not exhibit a bimodal distribution, but rather a peak at $v\approx260$\,km$\cdot$s$^{-1}$. This peak value is lower than those reported by \citet{2019A&A...626A..50D} for Be-type stars in the SMC cluster NGC 346 ($\sim$300\,km$\cdot$s$^{-1}$), by \citet{2022MNRAS.512.3331D} for Be-type stars in the LMC cluster 30 Dor ($\sim$310\,km$\cdot$s$^{-1}$), and by \citet{2023A&A...680A..32B} for Be-type stars in the SMC cluster NGC 330 ($\sim$400\,km$\cdot$s$^{-1}$). Generally, the rotational velocity of Be-type stars increase with decreasing metallicity \citep{2006A&A...452..273M}. In addition, the deconvolved rotational velocity distribution is affected by gravity darkening and binary interactions \citep{2016A&A...595A.132Z}.

We separately study the deconvolved $v$\,sin\,$i$ (true rotational velocity $v$) distribution for Be-type stars with different H$_\alpha$ emission lines and in different stellar environments. As shown in the left panel of Figure \ref{fig:b0},  the $v$ distribution of Be-shell stars exhibits an obvious bimodal distribution, while the $v$ distribution of Be-type stars with a single emission peak shows multiple peaks due to the small number of stars in that category. A K-S test indicates that Be-type stars in the field have similar $v$ distributions to those in OB associations and those in clusters individually (all $p>0.05$), but exhibit a highly significant difference from the combined sample of OB associations and clusters ($p<0.001$). This result suggests that there may be an intrinsic difference in the $v$ distribution between Be-type stars in high- and low-density star-forming regions (clusters and OB associations) and Be-type stars in the field. Detailed information is provided in Appendix~\ref{appd2} and the right panel of Figure~\ref{fig:b0}.

\subsection{Critical velocities of Be-type stars}

To test whether our Be-type stars are rotating at their critical velocities, we performed the following analysis. \citet{2022A&A...662A..66X} estimated the stellar parameters and elemental abundances of 33000 OBA-type stars in LAMOST DR6 using the Payne approach, based on LAMOST DR6 low-resolution spectra. We cross-matched the 479 Be-type stars with the catalog from \citet{2022A&A...662A..66X}, and obtained the effective temperature ($T_{\rm eff}$) and surface gravity (log\,$g$) for 105 stars with S/N$_g>$40. By using the solar-metallicity PARSEC isochrones \citep{2012MNRAS.427..127B} and $T_{\rm eff}$ and log\,$g$ of 105 stars, we derived their masses. Previous studies suggest that the $v$\,sin\,$i$ values of rapidly rotating stars are systematically underestimated due to gravity darkening and the absorption lines used to obtain $v$\,sin\,$i$ \citep{2005A&A...440..305F,2016A&A...595A.132Z,2004MNRAS.350..189T}. We followed the method of \citet{2022MNRAS.512.3331D} and used a cubic polynomial to obtain their corrected projected rotational velocities ($v$\,sin\,$i$$_{\rm correct}$).

Following the description of \citet{2013A&ARv..21...69R}, we calculated the critical rotational velocity ($v_{\rm crit}$), the rotational velocity of the Keplerian disk ($v_{\rm kd}$), and another critical rotational velocity ($v_{\rm c}$) of Be-type stars. Then, we obtained the $\gamma_{\rm e}$ values and the deconvolved $\gamma_{\rm e}$ distribution for the 105 stars. Further details are provided in Appendix~\ref{appd3}, and the stellar information is summarized in Table~\ref{tab:mc_ncrit_app1}.

Our results suggest that the corrected $v$\,sin\,$i$ distribution for these stars is consistent with $\gamma_{\rm cent}$$\sim$ 0.80. The deconvolved $\gamma_{\rm e}$ distribution reveals a bimodal feature: a minor peak at $\gamma_{\rm e}$$\sim$0.36, a primary peak at $\gamma_{\rm e}$$\sim$0.79, and a mean value of $\gamma_{\rm e}$$\sim$0.74 (see the left panel of Figure~\ref{fig:sidebyside}). The difference between the two is 0.06 (a relative deviation of about 8\%), which can be considered consistent after accounting for sample differences and uncertainties. The above results indicate that not all Be-type stars rotate at velocities close to the critical value \citep{2005A&A...440..305F}. Compared to the results from \citet{2016A&A...595A.132Z}, \citet{Bressert2010MNRAS.409L..54B}, and \citet{2021ApJ...921....5B} for the Galactic Be-type star sample, we obtain larger $\gamma_{\rm e}$ mean value. This is mainly due to the presence of a greater number of late-type Be-type stars in our sample \citep{2005ApJ...634..585C,2001A&A...368..912Y,2021ApJ...921....5B}. Previous studies report a dearth of Be-type stars with $\gamma_{\rm e}$$<$0.4 \citep{2016A&A...595A.132Z,2022MNRAS.512.3331D}. The bimodal distribution of deconvolved $\gamma_{\rm e}$ implies two different subpopulations: a slow component, possibly corresponding to Be-type stars that have recently undergone binary accretion but are not yet fully spun up; and a fast component, representing typical, fully evolved Be-type stars \citep{2020A&A...634A..51B}.

\subsection{Metallicity and rotational velocity}

\begin{figure*}[htp!]
    \centering
    \includegraphics[width=0.9\linewidth]{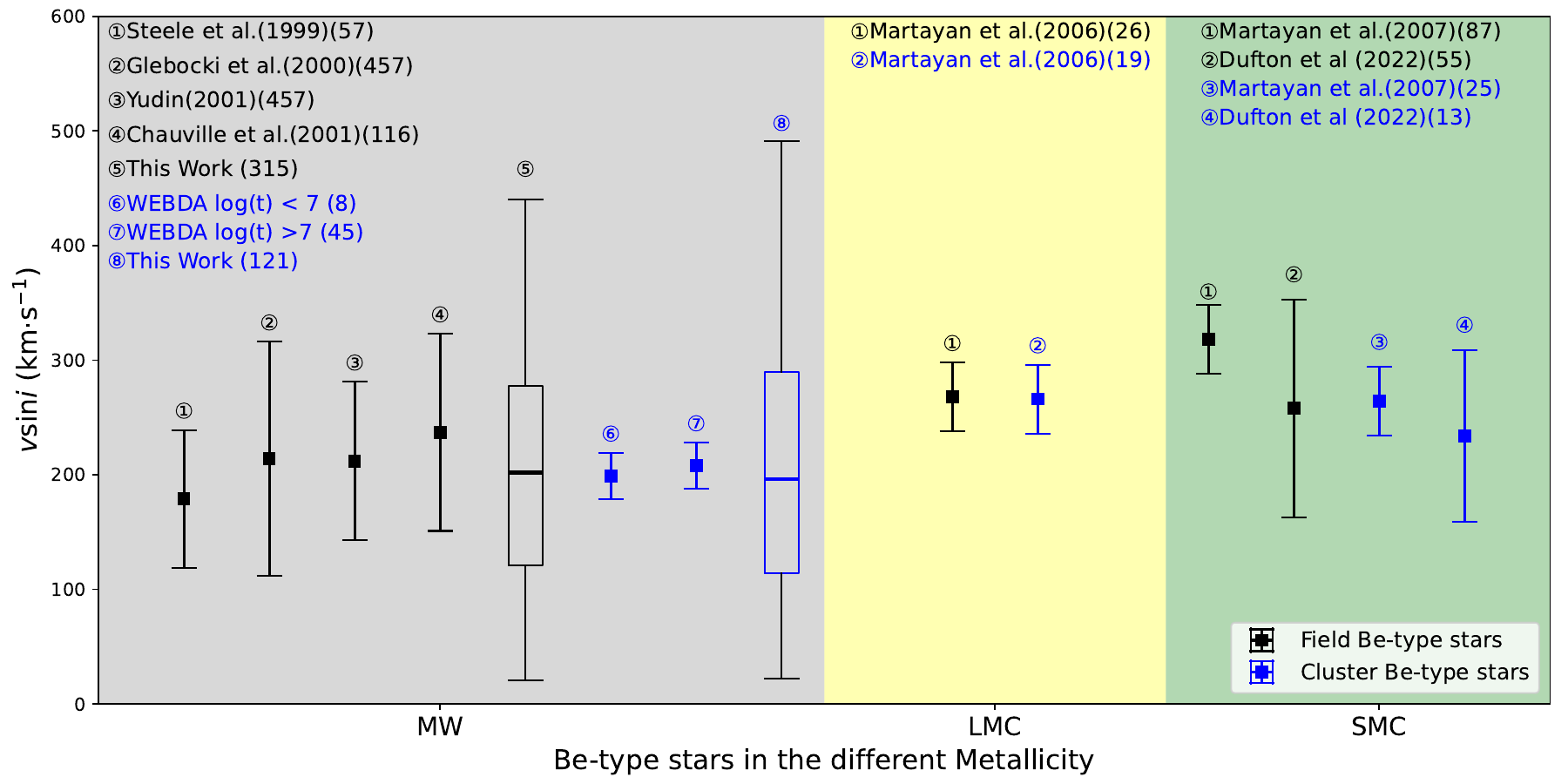}
    \caption{From left to right, the mean values of $v$\,sin$i$ for field and cluster Be-type stars in the MW, LMC, and SMC are compared. \textbf{The black and blue} squares represent the mean $v$\,sin\,$i$ of the field and cluster Be-type stars in the different literature, respectively. The numbers (\ding{192}, \ding{193}, \nodata) in each metallicity region correspond to specific labels. For example, \ding{192} refers to \citet{1992A&AS...95..437D}, where the number in parentheses (57) is the sample size. Be-type stars analyzed in this work are marked as a box.}
    \label{fig07}
\end{figure*}

Comparing rotational velocity distributions of Be-type stars in the Milky Way (MW) to those of the Large Magellanic Cloud (LMC) and the Small Magellanic Cloud (SMC), \citet{2006A&A...452..273M,2007A&A...462..683M} show that stars in the SMC have, on average, higher rotational velocities. To further study the effect of metallicity on the rotational velocity, we collected the rotational velocity data of Be-type stars in the MW, SMC, and LMC from the literature (MW \citep{Steele1999A&AS..137..147S,2000AcA....50..509G,2001A&A...368..912Y,2001A&A...378..861C}, LMC \citep{2006A&A...452..273M,2007A&A...462..683M}, and SMC \citep{2006A&A...452..273M,2007A&A...462..683M,2022MNRAS.512.3331D}). 

In Figure \ref{fig07}, we compare the mean $v$\,sin\,$i$ values of Be-type stars from this work and from the literature. From the distribution of the mean $v$\,sin\,$i$ of Be-type stars in the field and in clusters, we summarize the following results:

(i) For field Be-type stars: Our field Be-type stars have a similar mean value of $v$\,sin\,$i$ to MW Be-type stars reported in the literature. They have a lower average rotational velocity than those in the LMC and SMC.

(ii) For cluster Be-type stars: The mean $v\sin i$ of cluster Be stars in our sample is consistent with values reported for MW clusters in the literature.  However, the mean $v\sin i$ of cluster Be stars in the MW as a whole is lower than that of cluster Be stars in the LMC and SMC.

The field and cluster Be-type stars in our sample (in the MW) have lower rotational velocities than those in the LMC and SMC. This difference may be attributed to two factors: the field Be-type stars in lower-metallicity are more compact and, thus generally have higher rotational velocities \citep{2008A&A...478..467E}. They have lower mass-loss rates and therefore lose less angular momentum \citep{2007A&A...473..603M}.

\subsection{$v$\,sin\,$i$ and the distance of the double-peaked profiles of H$\alpha$}

\citet{1988A&A...189..147H} and \citet{1992A&AS...95..437D} found that the peak separations of H$_\beta$ and H$\alpha$ lines ($\Delta$V$_\beta$ and $\Delta$V$_\alpha$) of Be-type stars have the approximate relation $\rm \frac{\Delta V_\beta}{\Delta V_\alpha}$ =1.6-1.8 \footnote{$\Delta$V$_\beta$ and $\Delta$V$_\alpha$ are defined by $\Delta$V =(V$_{\rm R}$-V$_{\rm B}$), where V$_{\rm R}$ and V$_{\rm B}$ represent the radial velocities of H$_\beta$ and H$\alpha$ in the red and blue band for the double-peak profiles of Be-type stars, respectively}. Additionally, they noted that there is a correlation between $\Delta$V$_\alpha$ and $v$\,sin\,$i$. \citet{2023AN....34430022Z} revisited the distance between the peaks of H$\alpha$, H$_\beta$, and H$_\gamma$ emission lines by analyzing Be-type stars with high-resolution spectra. Considering the wavelength range of our spectra, we measured the $\Delta$V$_\alpha$ of 205 Be-type stars with double-peaked H$\alpha$ profiles. A Gaussian fit of the H$\alpha$ peak is shown in the right panel of Figure \ref{fig:sidebyside}.

\begin{figure}[htp!]
    \centering
    \includegraphics[width=9cm,height=7.5cm]{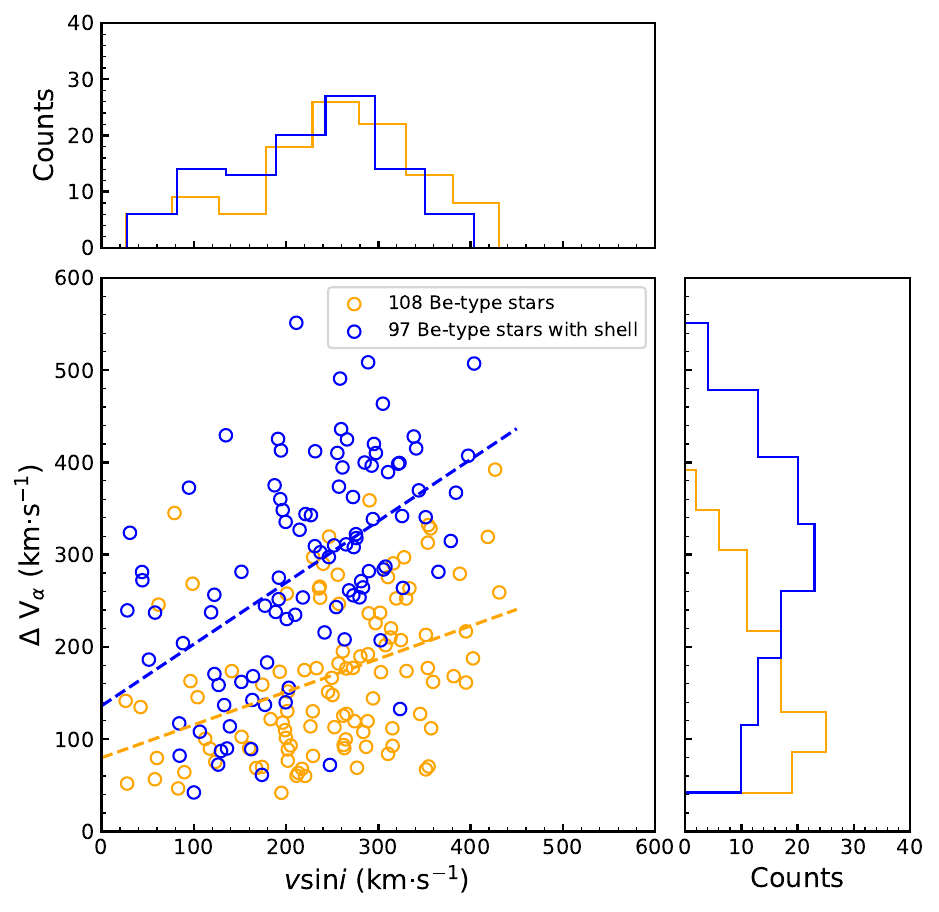}
    \caption{Distribution of $v$\,sin\,$i$ and $\Delta$V$_{\rm \alpha}$ for 108 Be-type stars (orange circles) and 97 Be-type stars with shell (blue circles). The blue and orange dashed lines represent the fitting results of the correlation between $v$\,sin\,$i$ and $\Delta$V$_{\rm \alpha}$. The different color histograms in the top and right panels correspond to the marks of the main panel.}
    \label{fig08}
\end{figure}

In Figure \ref{fig08}, we display the distribution of $v$\,sin\,$i$ and $\Delta$V$_\alpha$ for the 205 Be-type stars including 97 Be-type stars with shell. A positive correlation between $v$\,sin\,$i$ and $\Delta$V$_\alpha$ for the overall sample is consistent with previous studies \citep{1988A&A...189..147H,1992A&AS...95..437D}. For the 108 Be-type stars, the correlation is weak to moderate (Pearson $r$=0.397, $p$=2.09$\times$10$^{-5}$; Spearman $\rho$=0.433, $p$=2.88$\times$10$^{-6}$), while for the 97 shell stars it is stronger (Pearson $r$=0.514, $p$=4.63$\times$10$^{-8}$; Spearman $\rho$=0.533, $p$=1.13$\times$10$^{-8}$). Our results further reveal that Be-type shell stars tend to have larger $\Delta$V$_\alpha$ than non‑shell stars, which may be attributed to a more extended circumstellar envelope of ionized gas,  increasing the separation between the H$\alpha$ emission line peaks. The stronger correlation and steeper slope in Be-type shell stars are likely due to their nearly edge‑on orientation, where the measured H$\alpha$ more directly reflects the equatorial rotation velocity.

\section{Conclusions}\label{sec:Conclusions}

In this work, we measure the projected rotation velocities of 479 Be-type stars using the Fourier transform method, based on the LAMOST MRS spectra. Our main conclusions are as follows: 

(1) Our results suggest that the $v$\,sin\,$i$ values of Be-type stars can be derived using the FT method, based on the \ion{He}{1} lines at 4922, 5015, 5047, and 6678\,\AA.

(2) A K-S test indicates that Be-type stars with different H$\alpha$ emission line morphologies have different $v$\,sin\,$i$ distributions. The Be-type stars with double‑peak emission have larger $v$\,sin\,$i$ values than those with single-peak emission.

(3) Based on a literature search, we separate the Be-type stars into the field, OB associations, and open clusters. Clusters contain a higher proportion of rapidly rotating Be-type stars with $v$\,sin\,$i$$>$ 280\,km$\cdot$s$^{-1}$ than the field and OB associations, consistent with previous results \citep{2007AJ....133.1092W,2006A&A...452..273M,2006A&A...457..265D}. However, the K‑S test results for Be-type stars in clusters, OB associations, and the field indicate no significant differences in their $v$\,sin\,$i$ distributions (all $p>$0.05).

(4) The bimodal distribution of $v$\,sin\,$i$ for Be shell stars and field Be-type stars implies that there may be two different mechanisms producing them. However, the deconvolved $v$\,sin\,$i$ distribution of our entire Be-type star sample does not show a bimodal distribution, but rather a peak at $v\approx260$\,km$\cdot$s$^{-1}$. 

(5) Monte Carlo simulations of the 105 Be-type stars in our sample give a mean ratio of rotational to critical velocity $\gamma_{\rm e}$$\sim$0.8. The deconvolved $\gamma_{\rm e}$ distribution reveals a bimodal feature, with a minor peak at $\gamma_{\rm e}$$\sim$0.36, and a primary peak at $\gamma_{\rm e}$$\sim$0.79 (a mean value of $\gamma_{\rm e}$$\sim$0.74), which could indicate two distinct rotational subpopulations of Be-type stars. 

(6) We also calculated the Pearson correlation coefficients and Spearman's rank correlation coefficients between $v$\,sin\,$i$ and $\Delta$V$_\alpha$ for 205 Be-type stars and found that Be-type shell stars exhibit larger $\Delta$V$_\alpha$ values.

In our subsequent work, we will present the $v$\,sin\,$i$ values of all OB-type stars with LAMOST MRS spectra using the Fourier transform method, based on the \ion{He}{1} lines at 4922, 5015, 5047, and 6678\,\AA. We will also discuss the effects of metallicity, star formation conditions, and evolution on the distribution of $v$\,sin\,$i$ for OB stars.





We thank the anonymous referee for the helpful suggestions to improve this manuscript. This study is supported by the National Natural Science Foundation of China under grants Nos. 12403034, 12573026; the Natural Science Foundation of Hebei Province under grant No. A2024205031; Hebei Province Yan-zhao Golden Peak Talent Program (Postdoctoral Platform) for Key Talents under grant No. B2025003010, and the Science Foundation of Hebei Normal University (Nos. L2024B54, L2024B55, L2024B56, and L2026B51).  

The Guoshoujing Telescope, also known as the Large Sky Area Multi-Object Fiber Spectroscopic Telescope (LAMOST), is a National Major Scientific Project built by the Chinese Academy of Sciences. LAMOST is operated and managed by the National Astronomical Observatories, the Chinese Academy of Sciences. Research has made use of the SIMBAD database, operated at CDS, Strasbourg, France. This work has made use of the BeSS database, operated at LESIA, Observatoire de Meudon, France:(\url{http://basebe.obspm.f}).

\vspace{5mm}
Software:Topcat \citep{2005ASPC..347...29T}, Matplotlib \citep{2007CSE.....9...90H}, laspec \citep{2020ApJS..246....9Z,2021ApJS..256...14Z}

%






\appendix


\section{Synthetic spectra with different signal-to-noise ratio}
\label{appendb}

\renewcommand{\thefigure}{A1}
\begin{figure*}[htp!]
    \centering
    \includegraphics[width=1.0\linewidth]{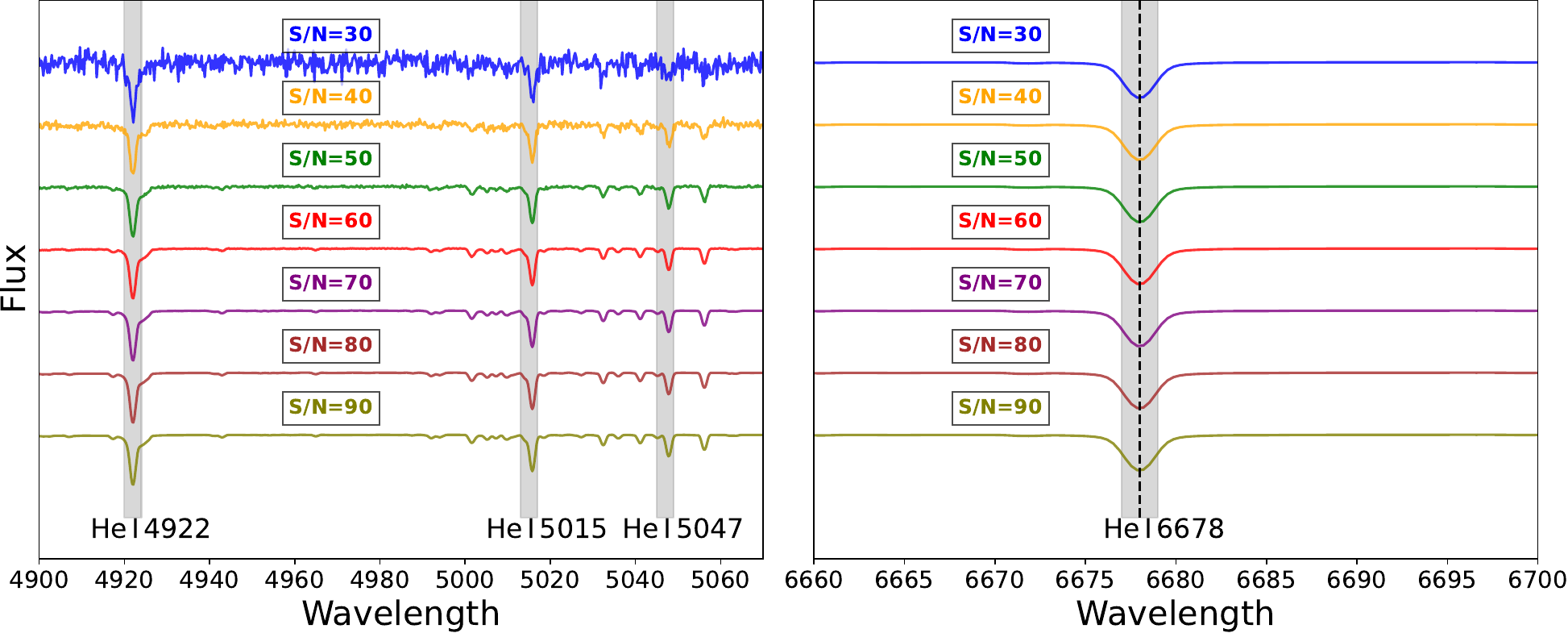}
    \caption{Synthetic spectra of a B-type star with $T_{\rm eff}$=20000\,K and log\,$g$=4.0\,dex at different signal-to-noise ratio (S/N) (30-90). Helium absorption lines (\ion{He}{1} $\lambda$4922, 5015, 5047, and 6678\,\AA) are also marked in the plots.}
    \label{fig:a}
\end{figure*}

To select Be-type stars with reliable signal-to-noise ratio (S/N) spectra, we calculate the synthetic spectra of a Be-type star with $T_{\rm eff}$=20000\,K and log\,$g$=4.0\,dex under various S/N from 30 to 90 with steps of 10. The synthetic spectra are obtained using the following website (\url{https://archive.stsci.edu/prepds/bosz/}) and noise is added to simulate a given S/N using Gaussian random values. In Figure~\ref{fig:a}, we show the synthetic spectra with different S/N. The spectral quality drops significantly when the S/N is lower than 40. Based on this test, we select the Be-type stars with spectral S/N$\geq$40 as our sample.

\section{deconvolved $v$\,sin\,$i$ for Be-type stars with different H$\alpha$ emission lines and in different stellar environments}\label{appd2}

Figure \ref{fig:b0} shows the distributions of probability density and cumulative fractions of the deconvolved $v$\,sin\,$i$ ($v$) for our Be-type stars with different H$\alpha$ emission lines and in different stellar environments. From the left panel of Figure \ref{fig:b0}, Be-shell stars show an obvious bimodal $v$ distribution, with peak velocities at 172\,km$\cdot$s$^{-1}$ and 266\,km$\cdot$s$^{-1}$. The $v$ distribution of field Be-type stars differs significantly from those of Be-type stars in both clusters ($p$=0.0235) and OB associations ($p$=0.024). The peak $v$ values of Be-type stars in clusters and OB associations are higher than those of field Be-type stars.

\renewcommand{\thefigure}{B1}
\begin{figure*}[htp!]
    \centering
    \includegraphics[width=1.0\linewidth]{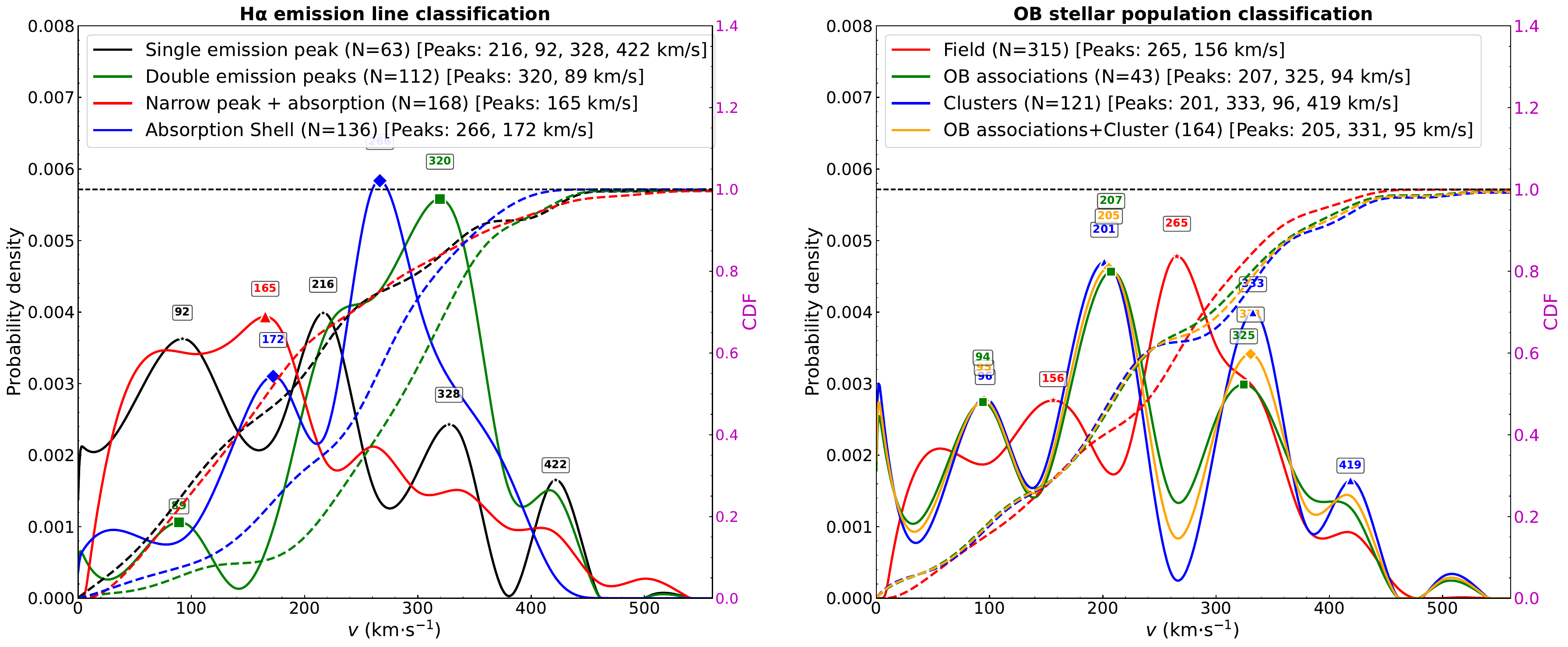}
    \caption{Left: probability density distributions (solid lines, left y-axis) and cumulative fractions (dashed lines, right y-axis) of deconvolved $v$\,sin\,$i$ ($v$) for Be-type stars with different H$_\alpha$ emission lines. The peak values for each subgroup are also indicated. Right: same as the left panel but for Be-type stars in clusters, the field, and OB associations.}
    \label{fig:b0}
\end{figure*}

\section{Calculation of critical velocities and $\gamma_{\rm e}$ distributions}\label{appd3}

\renewcommand{\thefigure}{C1}
\begin{figure*}[htbp]
    \centering
    \includegraphics[width=0.45\textwidth]{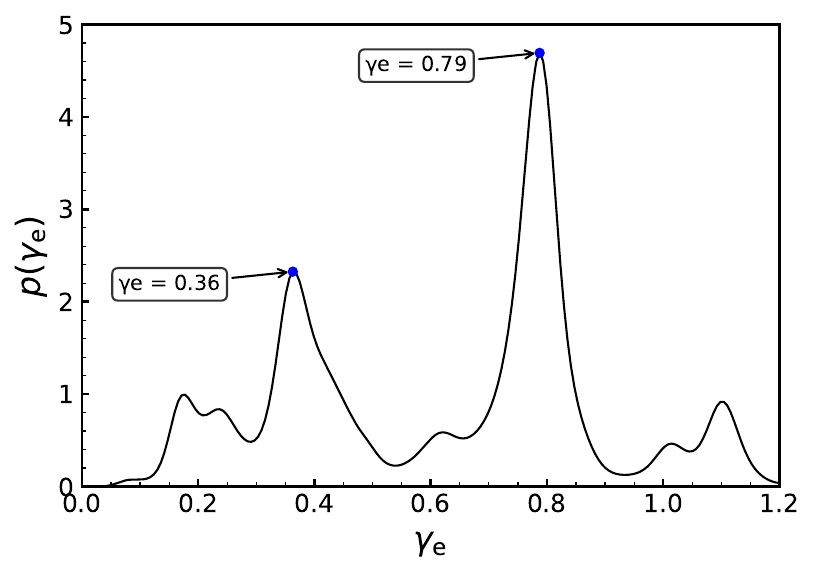}
    \hfill
    \includegraphics[width=0.45\textwidth]{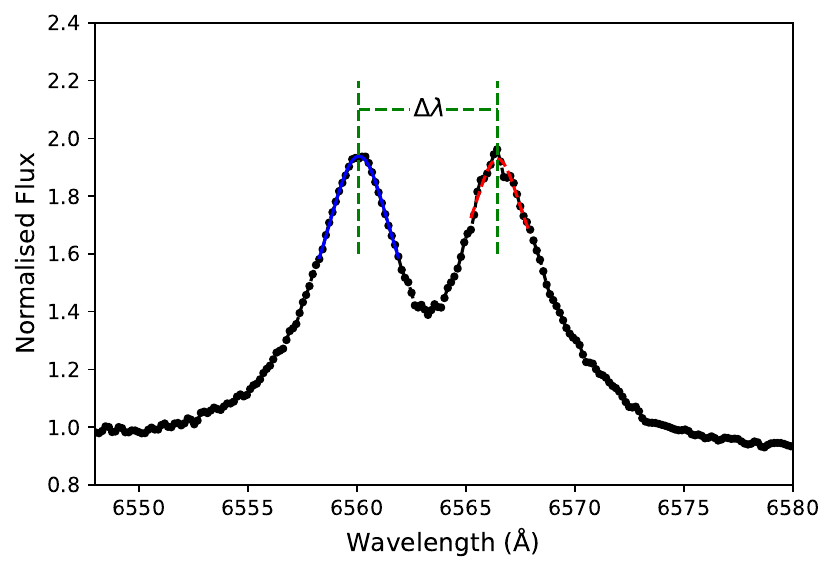}
    \caption{Left: the deconvolved $\gamma_{\rm e}$ distribution of the 105 stars in our sample. The peak values are also marked. Right: Gaussian fits of the H$\alpha$ double-peaked emission line for a Be-type star.}
    \label{fig:sidebyside}
\end{figure*}

Following the description of \citet{2013A&ARv..21...69R}, we calculated the critical rotation velocity ($v_{\rm crit}$) of Be-type stars when their equatorial velocity ($v_{\rm equa}$) reaches the rotational velocity of the Keplerian disk ($v_{\rm kd}$). We obtain the following relationships.

\begin{equation}
    v_{\rm kd} = \sqrt{\dfrac{GM}{R_{\rm e}}}
    \label{eq:kd}
\end{equation}

In equation~\eqref{eq:kd}, $M$ and $R_{\rm e}$ are the stellar mass and equatorial radius. For critical solid rotation, we have $R_e$=$\frac{2}{3}R_p$, where $R_p$ is the polar radius, and the $\frac{2}{3}$ factor arises from the oblateness \citep{2009pfer.book.....M}. Therefore, we can write the critical velocity ($v_{\rm crit}$):

\begin{equation}
    v_{\rm crit} = \sqrt{\dfrac{2GM}{3R_{\rm p}}} = \sqrt{\dfrac{2}{3}} (GM g_p)^{\frac{1}{4}}
    \label{eq:kd2}
\end{equation}

Where $g_p$ is the polar gravity, assuming that $R_p$ is relatively unaffected by rotation. When we equate the stellar surface gravity ($g$) with the polar gravity ($g_p$), we obtain another critical rotation velocity ($v_{c\rm}$):

\begin{equation}
    v_{\rm c} =\sqrt{\dfrac{2}{3}} (GM g)^{\frac{1}{4}}
    \label{eq:kd3}
\end{equation}

Based on the above definition, we obtain $v_{\rm c}\leq v_{\rm crit}\leq v_{\rm kd}$. Thus, $v_{\rm c}$ provides a lower limit for the critical velocity. Using the masses and surface gravities of 105 stars, we derived their $v_{\rm c}$.

In order to study the distribution of rotational velocities for our Be-type stars, we followed the method of \citet{2022MNRAS.512.3331D} and performed Monte Carlo simulations for a Gaussian distribution of rotational velocities.

{\large
\begin{equation}
    \begin{cases}
        p(\gamma_e) = \frac{1}{\sigma\sqrt{2\pi}} {\rm exp}\left(\frac{(\gamma_e-\gamma_{\rm cent})^2}{2\sigma^2}\right) \\[6pt]
        \gamma_e = \frac{v}{v_{\rm crit}} \\[6pt]
        F_{crit} = \frac{v {\rm sin} i}{v_{\rm crit}}
    \end{cases}
    \label{eq:group}
\end{equation}
}

In equation~\ref{eq:group}, we set $\sigma$ = 0.002 and selected seven $\gamma_{\rm cent}$ values ranging from 0.6 to 0.9 with a step size of 0.05. At the same time, we assumed a random $\sin\,i$ distribution to simulate the distribution of $v$\,sin\,$i$. Scaling according to the size of our sample, we obtained an estimate of the number of targets ($N_{\rm crit}$) whose expected rotational velocity lies below a certain fraction ($F_{\rm crit}$) of the critical velocity.

In Table~\ref{tab:mc_ncrit_app}, we list the results of Monte Carlo simulations for the number $N_{\mathrm{crit}}$ of our 105 stars with $v\sin i$ less than a fraction $F_{\mathrm{crit}}$ of the critical velocity ($v_{\mathrm{crit}}$). The observed and corrected $v$\,sin\,$i$ distributions for these stars are consistent with $\gamma_{\rm cent}$$\sim$0.85 and $\sim$ 0.80, respectively. The left panel of Figure~\ref{fig:sidebyside} shows the deconvolved probability distribution of $\gamma_{\rm e}$ from the corrected $v$\,sin\,$i$ for the above stars.

\renewcommand{\thetable}{C1}
\begin{table*}[ht!]
\centering
\caption{Monte Carlo simulations of the number $N_{\mathrm{crit}}$ of our 105 stars with $v\sin i$ less than a fraction $F_{\mathrm{crit}}$ of the critical velocity ($v_{\mathrm{crit}}$). The simulations assume a Gaussian distribution of equatorial rotation velocities (equation~\ref{eq:group}) with $\sigma = 0.02$ and seven $\gamma_{\rm cent}$ values ranging from 0.6 to 0.9 in steps of 0.05.}
\label{tab:mc_ncrit_app}
\begin{tabular}{lcccc}
\hline
$\gamma_{\mathrm{cent}}$ & \multicolumn{4}{c}{$F_{\mathrm{crit}}$} \\
\cline{2-5}   
& 0.5 & 0.4 & 0.3 & 0.2 \\
\hline
Observation     & 61 & 57 & 43 & 24 \\
correction      & 59 & 52 & 38 & 22 \\
\hline
$\gamma_{\mathrm{cent}}=0.60$ & $87.6\pm3.8$ & $70.1\pm4.8$ & $52.6\pm5.2$ & $35.1\pm4.9$ \\
$\gamma_{\mathrm{cent}}=0.65$ & $80.9\pm4.4$ & $64.8\pm5.0$ & $48.6\pm5.1$ & $32.4\pm4.7$ \\
$\gamma_{\mathrm{cent}}=0.70$ & $75.1\pm4.6$ & $60.1\pm5.1$ & $45.1\pm5.1$ & $30.0\pm4.6$ \\
$\gamma_{\mathrm{cent}}=0.75$ & $70.1\pm4.8$ & $56.0\pm5.1$ & $42.1\pm5.0$ & $28.1\pm4.5$ \\
$\gamma_{\mathrm{cent}}=0.80$ & $65.7\pm4.9$ & $52.6\pm5.1$ & $39.4\pm4.9$ & $26.3\pm4.4$ \\
$\gamma_{\mathrm{cent}}=0.85$ & $61.8\pm5.1$ & $49.4\pm5.1$ & $37.0\pm4.9$ & $24.7\pm4.3$ \\
$\gamma_{\mathrm{cent}}=0.90$ & $58.4\pm5.1$ & $46.7\pm5.1$ & $35.0\pm4.8$ & $23.3\pm4.3$ \\
\hline
\end{tabular}
\end{table*}

\renewcommand{\thetable}{C2}
\begin{table*}[ht!]
\centering
\caption{The information of 105 stars in our Be-type stars.}
\label{tab:mc_ncrit_app1}
\begin{tabular}{cccccccc}
\hline
Designation &$T_{\rm eff}$$^a$&log\,$g$$^a$& Mass& $v$\,sin\,$i_{\text{This work}}^b$&$v_{c\rm }$$^c$ &$v$\,sin\,$i$$_{\rm correct}^d$\\ 
(LAMOST)&(K)&(dex)&(M$_\odot$)&km$\cdot$s$^{-1}$&km$\cdot$s$^{-1}$&km$\cdot$s$^{-1}$\\
\hline
J014915.82+535821.8 & $16017 \pm 355 $ & $4.18\pm 0.09$ & 4.6 & 359.2 & 450.5 & 377.9\\
J020144.70+564904.0 & $17083 \pm 269 $ & $4.18\pm 0.07$ & 5.0 & 134.7 & 460.4 & 144.4\\
J020728.03+601610.7 & $17319 \pm 1399$ & $2.47\pm 0.24$ & 29.9& 211.0 & 266.1 & 221.5\\
J020857.51+562108.5 & $23195 \pm 436 $ & $4.70\pm 0.10$ & 8.0 & 26.1  & 696.8 & 25.2\\
J021300.50+550639.5 & $22635 \pm 1350$ & $4.07\pm 0.32$ & 8.9 & 288.7 & 496.3 & 300.7\\
J021437.15+560432.6 & $23771 \pm 3453$ & $3.43\pm 0.70$ & 13.6& 388.3 & 373.7 & 412.3\\
\nodata&\nodata&\nodata&\nodata&\nodata&\nodata\\
\nodata&\nodata&\nodata&\nodata&\nodata&\nodata\\
\nodata&\nodata&\nodata&\nodata&\nodata&\nodata\\
\hline
\end{tabular}
\begin{tablenotes}
\item[] Notes: 
\item[] $^a$: The $T_{\rm eff}$ and log\,$g$ from \citet{2022A&A...662A..66X}.
\item[] $^b$: The $v$\,sin\,$i$ of Be-type stars obtained using FT in this work.
\item[] $^c$: The critical velocities of Be-type stars using their stellar parameters (see equation~\ref{eq:kd3}).
\item[] $^d$: The corrected $v$\,sin\,$i$ of Be-type stars using a cubic polynomial from \citep{2022MNRAS.512.3331D}.
\item[] A full version of this FITS format table is available electronically. 
\end{tablenotes}
\end{table*}





\bibliography{sample631}{}
\bibliographystyle{aasjournal}



\end{document}